\documentclass[aps,prl,twocolumn,superscriptaddress]{revtex4-2}

\usepackage{amsmath,amssymb,bm}
\usepackage{graphicx}
\usepackage{microtype}
\usepackage{xcolor}
\definecolor{LinkBlue}{RGB}{0,70,170}
\usepackage[colorlinks=true,citecolor=LinkBlue,linkcolor=LinkBlue,urlcolor=LinkBlue]{hyperref}

\newcommand{\Tr}{\operatorname{Tr}}
\newcommand{\Var}{\operatorname{Var}}
\newcommand{\ee}{\mathrm{e}}

\begin{document}

\title{Entanglement Embezzlement from Diffusive Hydrodynamics}

\author{Shi-Xin Zhang}
\email{shixinzhang@iphy.ac.cn}
\affiliation{Institute of Physics, Chinese Academy of Sciences, Beijing 100190, China}

\author{Shuo Liu}
\email{sl6097@princeton.edu}
\affiliation{Department of Physics, Princeton University, Princeton, New Jersey 08544, USA}

\author{Yu-Qin Chen}
\email{yqchen@gscaep.ac.cn}
\affiliation{Graduate School of China Academy of Engineering Physics, Beijing 100193, China}

\date{September 21, 2026}

\begin{abstract}
Entanglement embezzlement asks how much entanglement can be borrowed from a many-body state by local operations and classical communication (LOCC) while returning that state with a small error. We uncover a conservation-law mechanism that redistributes the dominant probability mass of the Schmidt spectrum in the logarithmic Schmidt-rank coordinate and thereby controls this operational resource. For typical random pure states at fixed U(1) charge, we prove a finite-error conversion law: away from half filling, the charge bias converts $O(\sqrt L)$ charge fluctuations into $O(\sqrt L)$ borrowable entanglement, whereas particle--hole symmetry at half filling removes this contribution entirely. We then show how the resource develops dynamically in charge-conserving random circuits. Combining hydrodynamic analysis with large-scale replica tensor-network calculations, we find that diffusion broadens the operationally relevant distribution in the logarithmic Schmidt-rank coordinate on the scale $t^{1/4}$ and increases the amount of entanglement that can be borrowed. Charge transport therefore continues to reorganize the entanglement spectrum and activate embezzlement after the leading volume-law entropy has saturated.
\end{abstract}

\maketitle

\textit{Introduction.---} Quantum thermalization is conventionally characterized by the rapid generation of macroscopic entanglement and the erasure of local memory~\cite{deutsch1991quantum,srednicki1994chaos,dalessio2016thermalization,eisert2015nonequilibrium,vidmar2017eigenstates,kaufman2016thermalization,islam2015measuring,zhang2026entangledgrowth}. However, a many-body state need not reach full global equilibrium on a single timescale. For instance, in one-dimensional chaotic systems, the leading volume-law entanglement entropy grows ballistically~\cite{calabrese2005evolution,calabrese2007quench,kim2013ballistic,mezei2017spreading,nahum2017quantum}, whereas conserved densities relax much more slowly through diffusive hydrodynamics~\cite{rakovszky2018diffusive,khemani2018}. This temporal separation implies that, long after the primary entanglement entropy has saturated, the state's internal entanglement structure can continue to evolve, as reflected in higher R\'enyi entropies and other nonlinear resources~\cite{rakovszky2019,zhou2020diffusive,liu2026participation,aditya2026coherence,xiao2026diffusive,pawlik2026restricted}. Capturing this late-time evolution requires looking beyond individual entropy measures to the full entanglement spectrum, which records the redistribution of probability weight among Schmidt ranks~\cite{li2008entanglement,calabrese2008entanglement,torre2026nonlocalmagic,lubkin1978entropy,page1993average,nadal2011statistical}. This raises a natural operational question: after the extensive entropy has formed, can this slow spectral reorganization enable a quantum-information task that the saturated entropy alone cannot predict?

As a prime candidate for such a spectrum-sensitive task, entanglement embezzlement starts from a shared catalyst $|\psi\rangle_{A_{\rm c}B_{\rm c}}$ and unentangled target registers. Alice and Bob use local operations and classical communication (LOCC) to approximate $|\psi\rangle_{A_{\rm c}B_{\rm c}}\otimes|\Phi_m\rangle_{A_{\rm t}B_{\rm t}}$, where $|\Phi_m\rangle=m^{-1/2}\sum_{j=1}^m|j\rangle_{A_{\rm t}}|j\rangle_{B_{\rm t}}$ is an $m$-dimensional maximally entangled state, while returning the catalyst almost unchanged~\cite{bennett1996concentrating,jonathan1999entanglement,vanDam2003embezzling,cleve2017perfect,zanoni2024complete,horodecki2009quantum,chitambar2019resource}. The feasibility of this conversion is governed by majorization~\cite{nielsen1999conditions}, and the rank-octave criterion identifies universal embezzling families~\cite{zanoni2024complete,sierant2026nonlocalmagic}. Thus the task depends on the fine-grained distribution of Schmidt probabilities across ranks rather than only on the total entanglement~\cite{zhang2026magicbarrier,zhang2026schmidtscales,aditya2026nonlocalmagic,li2026nonlocalmagic}. Critical states, Gaussian fermions, structured many-body transformations, and chaotic circuits furnish realizations of Schmidt-level distributions~\cite{vanLuijk2025critical,kera2025gaussian,schwartzman2024complexity,karjula2026embezzlement}. While these works establish the existence of such spectrum-sensitive resources in specific states or models, the central question remains how this resource emerges dynamically in a generic thermalizing many-body system, and which physical process controls its finite-error borrowing capacity.

Here we answer this question by showing that the dynamical emergence of entanglement embezzlement is controlled by conserved-charge fluctuations. The mechanism comes from the interplay between a global conservation law and the subsystem boundary. In a system with fixed total charge, the subsystem charge $Q_A$ fluctuates. Away from half filling, an effective charge bias $\mu$ converts a fluctuation $\delta Q_A$ into a displacement $\mu\delta Q_A$ of the entanglement energies, thereby spreading the probability weight of the Schmidt spectrum in the logarithmic Schmidt-rank coordinate. Sector-weight fluctuations, often studied in symmetry-resolved entanglement~\cite{goldstein2018symmetry,xavier2018equipartition}, thus acquire a direct operational meaning: they determine how far LOCC can shift the ordered Schmidt spectrum while leaving the catalyst nearly unchanged. At half filling, particle--hole symmetry enforces $\mu=0$ and removes the leading charge-induced spectral broadening.

To formalize this picture, we develop a theory that links charge fluctuations to finite-error LOCC conversion. For equilibrium configurations modeled by fixed-charge random states, it gives the optimal fidelity for borrowing an $m$-dimensional maximally entangled state, equivalently $\log_2m$ ebits. To resolve the nonequilibrium timescales in charge-conserving random circuits, we combine hydrodynamic analysis with a charge-adapted two-replica tensor-network calculation and show how slow diffusive relaxation translates into the delayed reorganization of the entanglement spectrum as well as the growth of borrowable entanglement~\footnote{See Supplemental Material for proofs, methods, and additional numerical results.}.

\textit{Operational entanglement borrowing.---} Let $|\psi\rangle_{A_{\rm c}B_{\rm c}}$ be the many-body catalyst across a spatial cut. The target registers begin in the product state $|0\rangle_{A_{\rm t}}|0\rangle_{B_{\rm t}}$, and Alice and Bob use LOCC to approximate $|\psi\rangle\otimes|\Phi_m\rangle$ for an integer $m\geq1$; the target has Schmidt rank $m$ and contains $\log_2m$ ebits. Writing $F(\rho,\sigma)=\|\sqrt\rho\sqrt\sigma\|_1$ for root fidelity, let $F_m^\star$ denote the maximum root fidelity between the complete LOCC output and this ideal catalyst--target state. We define
\begin{align}
 E_{\rm borrow}(\epsilon)=\max_{m\in\mathbb N}\{\log_2m:F_m^\star\geq\sqrt{1-\epsilon^2}\}.
 \label{eq:task}
\end{align}
Thus $E_{\rm borrow}$ is the number of ebits that can be borrowed at global error tolerance $\epsilon$. Because the fidelity compares the entire output, it simultaneously penalizes catalyst disturbance, target infidelity, and residual catalyst--target correlations.

Varying the target dimension $m$ gives a family of conversion fidelities. At a fixed error tolerance, $E_{\rm borrow}(\epsilon)$ selects the largest target whose fidelity remains above the chosen threshold. This retains information that a single entanglement entropy discards: two states with similar entropy can distribute their Schmidt probabilities very differently across ranks and can therefore support different borrowing capacities. The family $F_m^\star$ is consequently well suited to the slow spectral reorganization studied below, while $E_{\rm borrow}$ reduces it to one directly interpretable number.

\begin{figure}[!t]
 \includegraphics[width=\columnwidth]{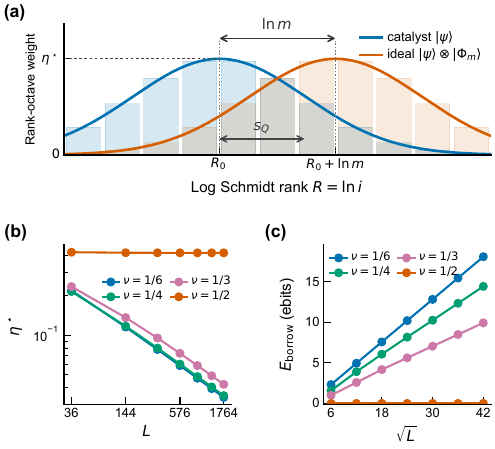}
 \caption{\label{fig:equilibrium}Logarithmic Schmidt-rank-coordinate mechanism and fixed-charge random states at $L_A=L/3$. (a) Gaussian limiting schematic of the mass per rank octave versus $R=\ln i$. Charge fluctuations produce a broad envelope of characteristic width $s_Q$; tensoring the ideal $m$-dimensional target translates it by exactly $\ln m$, while its maximum octave mass is $\eta^\star$. (b) Exact $\eta^\star$ is plotted against $L$ on a log--log scale; it decays at $\nu=1/6,1/4,1/3$ and remains finite at $\nu=1/2$. (c) At global error $\epsilon=0.5$, $E_{\rm borrow}$ grows linearly with $\sqrt L$ away from half filling but has no growing half-filled contribution. Panels (b) and (c) evaluate the exact finite-$L$ ensemble-averaged marginal; the theorem transfers their asymptotic laws to typical fixed-charge pure states.}
\end{figure}

Let $p=(p_1,\ldots,p_d)$ be the decreasing vector of squared Schmidt coefficients of $|\psi\rangle$. In the common dimension $M=md$, the source is $p$ padded by zeros and the ideal catalyst--target state has Schmidt vector $v=p\otimes(1/m,\ldots,1/m)$. Nielsen's majorization criterion and the reduction of arbitrary mixed outputs to deterministic pure outputs for a pure target~\cite{nielsen1999conditions,vidal2000approximate} imply that the relevant output Schmidt vectors obey the cumulative inequalities $r\succ p$, meaning $\sum_{i=1}^k r_i\geq\sum_{i=1}^k p_i$ for $k<M$, with equality at $M$. The optimization is therefore
\begin{align}
 F_m^\star=\max_{r\succ p}\sum_{i=1}^{M}\sqrt{r_iv_i}.
 \label{eq:finite-fidelity-optimum}
\end{align}
We obtain this deterministic and global optimum by applying the pool-adjacent-violators algorithm to resolve the cumulative majorization constraints. While mathematically equivalent to the block solution in Ref.~\cite{vidal2000approximate}, our implementation is specifically optimized for the highly degenerate fixed-charge spectra by storing only distinct eigenvalues and their multiplicities~\cite{Note1}. Figure~\ref{fig:equilibrium}(a) gives the corresponding Schmidt weight distribution in the logarithmic Schmidt-rank coordinate. Tensoring the target replaces each $p_i$ by $m$ copies of $p_i/m$, translating the entire probability profile by exactly $\ln m$ on the logarithmic Schmidt-rank coordinate $R=\ln i$. The established octave statistic~\cite{sierant2026nonlocalmagic,zhang2026schmidtscales}
\begin{equation}
 \eta^\star=\max_{k\geq1}\sum_{i=k}^{\min(2k-1,d)}p_i,
 \label{eq:eta}
\end{equation}
is the largest mass in any fixed-width interval $\Delta R=\ln2$. Its vanishing, $\eta^\star\to0$ when $L\rightarrow \infty$, is precisely the established criterion for a family of states to be a universal entanglement embezzler under unrestricted LOCC where no fixed-width rank octave retains finite probability. Going beyond this asymptotic criterion, $E_{\rm borrow}(\epsilon)$ measures how large a translation can be tolerated at finite error~\cite{vidal2000monotones,klimesh2007catalytic,zanoni2024complete}.

\textit{Fixed-charge random states.---} We first study the Haar-random fixed-charge state, which is the equilibrium model for chaotic particle-number-conserving dynamics and provides the static benchmark for the circuit evolution below. Consider a pure state in the $N=\nu L$ charge sector and a region of size $L_A=fL<L/2$. The ensemble-averaged marginal is explicitly block flat,
\begin{align}
 \overline\rho_A&=\bigoplus_q\lambda_q\mathbb I_{d_A(q)},\qquad d_A(q)=\binom{L_A}{q},\notag\\
 \lambda_q&=\frac{\binom{L-L_A}{N-q}}{\binom{L}{N}}.
 \label{eq:block-spectrum}
\end{align}
Here $q$ is the charge in $A$, while the environment carries $N-q$. Thus the charge-$q$ block contains $d_A(q)$ identical eigenvalues $\lambda_q$ and has total probability mass $p_q=d_A(q)\lambda_q$. This distinction between eigenvalue, degeneracy, and sector mass is the basic spectral picture used below. The masses $p_q$ form a hypergeometric distribution whose charge variance is $\sigma_Q^2\equiv\Var(Q_A)=f(1-f)\nu(1-\nu)L+O(1)$. Writing the charge displacement from the center as $\delta q=q-\nu L_A$, the eigenvalue varies across the typical window as
\begin{equation}
 -\ln\lambda_q=c_L+\mu\delta q+O(1),\qquad \mu=\ln\frac{1-\nu}{\nu}.
 \label{eq:chargewidth}
\end{equation}
 Here $c_L$ is a $q$-independent offset that sets the center of the entanglement-energy profile. Equation~\eqref{eq:chargewidth} has a direct meaning in the logarithmic Schmidt-rank coordinate. Across the typical charge sectors, moving by one sector changes both $-\ln\lambda_q$ and the logarithm of the cumulative rank by $|\mu|$, up to subleading offsets. The Gaussian charge distribution of width $\sigma_Q$ therefore becomes a Gaussian probability envelope in the logarithmic Schmidt-rank coordinate with width $s_Q(L)\equiv|\mu|\sigma_Q$, as drawn in Fig.~\ref{fig:equilibrium}(a). A local central-limit analysis then gives $\eta^\star=\Theta(L^{-1/2})$, as in panel (b)~\cite{Note1}.

The same width determines the best attainable fidelity for borrowing $\log_2m$ ebits. Writing the target shift in units of this width as $x\equiv\ln m/s_Q(L)$, we prove that, for $x$ held fixed as $L\to\infty$, the optimal fidelity obeys
\begin{equation}
 F_m^\star\longrightarrow\int_{-\infty}^{\infty}\!dz\sqrt{\phi(z)\phi(z+x)}
 =\ee^{-x^2/8}.
 \label{eq:normal-limit}
\end{equation}
Here $\phi$ is the standard normal density. The equation directly expresses the picture in Fig.~\ref{fig:equilibrium}(a): charge fluctuations produce the Gaussian catalyst envelope, the ideal target translates it by $\ln m$, and the square-root sum in Eq.~\eqref{eq:finite-fidelity-optimum} becomes the overlap of the two envelopes~\cite{Note1}. The geometric picture is therefore simple: a broader probability envelope makes the same translation $\ln m$ occupy a smaller fraction of its width. The two envelopes overlap more strongly, which raises the conversion fidelity and allows larger borrowable entanglement.

To elevate this result from the ensemble average to a typical instance, we promote the analytic mean spectrum to the physical Schmidt spectrum of a typical fixed-charge pure state. A blockwise Dirichlet--Wishart calculation gives an exponentially small trace distance between the two at every fixed unequal cut $f<1/2$~\cite{popescu2006entanglement,goldstein2006canonical,reimann2007typicality,zyczkowski2001induced}. Inverting Eq.~\eqref{eq:normal-limit} yields
\begin{align}
 E_{\rm borrow}(\epsilon)&=\kappa_\epsilon|\mu|\sqrt{f(1-f)\nu(1-\nu)L}+o(\sqrt L),\notag\\
 \kappa_\epsilon&=\frac{2\sqrt{-\ln(1-\epsilon^2)}}{\ln2}.
 \label{eq:borrowing-law}
\end{align}
This is the static operational law in Fig.~\ref{fig:equilibrium}(c): the ordinary $O(\sqrt L)$ charge window becomes an $O(\sqrt L)$ embezzlement resource because each transported charge moves the typical Schmidt mass by $\mu$ in the logarithmic Schmidt-rank coordinate.

At half filling, particle--hole symmetry explicitly sets the effective charge bias to $\mu = 0$. Without this linear mapping, the $O(\sqrt{L})$ charge fluctuations fail to translate into a macroscopic spread in the logarithmic Schmidt-rank coordinate. Instead, the probability mass remains tightly concentrated within an $O(1)$ width in the logarithmic Schmidt-rank coordinate. This extreme spectral concentration implies that a single factor-two rank window captures a finite fraction of the total probability, immediately yielding $\eta^\star = \Theta(1)$. The lack of a broad spectral envelope restricts the tolerable finite-error translation, capping $E_{\rm borrow}(\epsilon)$ at $O(1)$ independent of system size. The system thus presents an interesting contrast: while the half-filled state harbors the maximum volume-law entanglement entropy, its lack of charge bias gives it the weakest growing embezzlement resource. In sum, we demonstrate that typical random pure states at fixed $U(1)$ charge are universal entanglement embezzlers except at half filling.

\textit{Diffusive logarithmic Schmidt-rank broadening.---} We now follow how this fixed-charge spectrum forms dynamically. Starting from the period-6 repeating product state $|100000\,100000\cdots\rangle$ at filling $\nu=1/6$, we evolve an open chain with independent random U(1)-conserving two-qubit gates in a brick-wall geometry and take $A$ to be the leftmost $L/3$ sites. Such symmetric random circuits display charge-sector-dependent relaxation and hydrodynamics~\cite{liu2024symmetry,nahum2018operator,vonKeyserlingk2018operator,zhou2019emergent,mcculloch2023full, xiao2026nonstabilizernessmpembaeffects}. After averaging the random gates, the computational-basis occupations follow a symmetric exclusion process: hard-core particles hop diffusively while the total charge remains fixed. With diffusion constant $D_{\rm diff}$, nonuniversal amplitude $a$, and charge bias $\mu=\ln[(1-\nu)/\nu]=\ln5$, we define the width of the spectrum's probability distribution in the logarithmic Schmidt-rank coordinate, $s_Q(t,L)$, through the pre-saturation full counting statistics,
\begin{align}
 \Var Q_A(t)&\sim a\sqrt{D_{\rm diff}t},\notag\\
 s_Q(t,L)&\equiv|\mu|\sqrt{\Var Q_A(t)}\propto t^{1/4}.
 \label{eq:dynamics}
\end{align}
The fourth root has a direct spectral meaning: diffusion grows the charge variance as $\sqrt t$. At finite size, $\Var Q_A(t,L)=L\mathcal V(t/L^2)+o(L)$, where $\mathcal V$ is the finite-size scaling function, predicts a common diffusive clock $t/L^2$ for $s_Q(t,L)/\sqrt L$.

This dynamical prediction is the time-dependent continuation of Eq.~\eqref{eq:chargewidth}. In the final state, $s_Q(L)\sim\sqrt L$; during the hydrodynamic window until $t\sim O(L^2)$, the width grows as $s_Q(t,L)\sim t^{1/4}$. The normal-limit conversion law predicts $E_{\rm borrow}\propto s_Q(t,L)$ whenever the charge envelope is smooth~\cite{karjula2026embezzlement,aditya2026nonlocalmagic}.

\begin{figure}[!t]
 \includegraphics[width=\columnwidth]{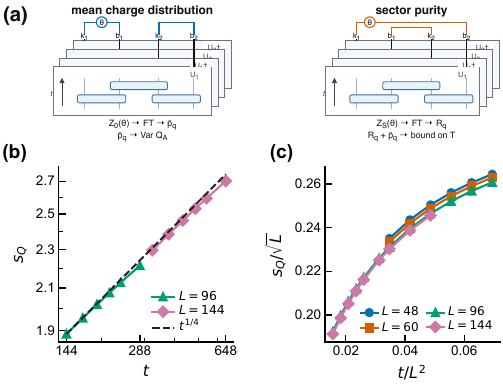}
 \caption{\label{fig:hydrodynamics}Diffusion redistributes probability in the logarithmic Schmidt-rank coordinate. (a) Identity and swap closures of the same Haar-averaged four-layer fold yield $\bar p_q$ and $R_q$, respectively; together they bound the structure omitted by charge-block flattening. Here FT denotes the discrete Fourier transform. (b) The charge-generated width $s_Q(t,L)=|\mu|\sqrt{\Var Q_A}$ is plotted against $t$ on a log--log scale for $L=96,144$ and follows the $t^{1/4}$ guide, with descriptive slopes $0.233,0.236$. (c) The scaled width $s_Q(t,L)/\sqrt L$ is plotted against $t/L^2$ on linear scales, collapsing the data for $L=48,60,96,144$.}
\end{figure}

We access the large-system regime by analytically averaging each random gate and evolving the deterministic two-copy tensor network $\mathcal M_2(t)=\mathbb E_c[\rho_c(t)^{\otimes2}]$ as a symmetry-adapted matrix-product state with a two-site time-evolving block-decimation update implemented in TeNPy~\cite{weingarten1978,collins2006,vidal2003,orus2008,singh2010,hauschild2018tenpy}. Figure~\ref{fig:hydrodynamics}(a) shows the two top-boundary contractions used to obtain the mean charge distribution and sector purity. With the ordinary identity closure and a phase $\ee^{i\theta Q_A}$ on one replica,
\begin{align}
 Z_0(\theta)&=\mathbb E_c\Tr(\ee^{i\theta Q_A}\rho_{A,c})
 =\sum_q\ee^{i\theta q}\bar p_q,
 &\bar p_q&=\mathbb E_c p_{q,c}.
 \label{eq:chargegenerator}
\end{align}
Because $q=0,\ldots,L_A$ is integer valued, an exact discrete Fourier transform of $Z_0$ gives the mean charge distribution $\bar p_q$. It fixes $\Var Q_A$, then $s_Q(t,L)=|\mu|\sqrt{\Var Q_A}$, and also the block-flat proxy $\overline{\widetilde\rho}_A=\bigoplus_q[\bar p_q/d_A(q)]\mathbb I_{d_A(q)}$. Replacing the identity closure on $A$ by the replica swap gives
\begin{align}
 Z_{\mathbb S}(\theta)&=\mathbb E_c\Tr(\ee^{i\theta Q_A}\rho_{A,c}^2)\notag\\
 &=\sum_q\ee^{i\theta q}R_q,
 \qquad R_q=\mathbb E_c\Tr\rho_{A,q,c}^2,
 \label{eq:puritygenerator}
\end{align}
where $\rho_{A,q,c}=\Pi_q\rho_{A,c}\Pi_q$ is the unnormalized charge-$q$ block. Define the true reduced state $\rho_{A,c}=\bigoplus_q\rho_{A,q,c}$ and its block-flattened version $\widetilde\rho_{A,c}=\bigoplus_q[p_{q,c}/d_A(q)]\mathbb I_{d_A(q)}$. Their trace distance $T(\rho_{A,c},\widetilde\rho_{A,c})$ measures the spectral information lost by erasing structure within each charge sector. The purity and $\bar p_q$ yield the conservative two-copy bound
\begin{equation}
 \mathbb E_c T(\rho_{A,c},\widetilde\rho_{A,c})
 \leq\frac12\sum_q\sqrt{d_A(q)R_q-\bar p_q^2}.
 \label{eq:mainsectorbound}
\end{equation}
The first readout therefore constructs the charge envelope used by the majorization calculation, while $R_q$ and $\bar p_q$ bound, in ensemble mean, the error incurred by flattening each sector~\cite{Note1}.

Figure~\ref{fig:hydrodynamics}(b) resolves the fourth-root regime at fixed large sizes: the descriptive exponents for $L=96$ and $144$ are already close to $1/4$ in the small-$t/L^2$ window. We do not fit the smaller sizes because their ballistic entanglement and diffusive charge-relaxation times are too close to leave a clean temporal window. Panel (c) supplies the independent finite-size scaling test, with four lengths organizing on the same diffusive clock and the two largest sizes agreeing at the sub-percent level in the same small-$t/L^2$ window. The agreement of the fixed-size slopes and finite-size collapse places the motion of the charge-generated spectral envelope on the hydrodynamic clock, parametrically later than leading entropy growth.

\begin{figure}[!t]
 \includegraphics[width=\columnwidth]{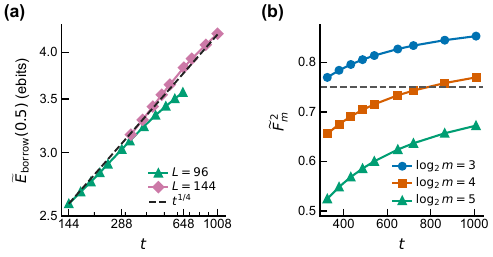}
 \caption{\label{fig:operational}Diffusive logarithmic Schmidt-rank broadening activates embezzlement. (a) The continuously interpolated yield $\widetilde E_{\rm borrow}(0.5)$ is plotted against $t$ on a log--log scale and follows the fourth-root guide in the intermediate window. The green $L=96$ curve bends away once charge fluctuations approach finite-size saturation, while $L=144$ exceeds four ebits. (b) The best global squared fidelity at $L=144$ is plotted against $t$ on linear scales. The dashed line marks the $0.75$ threshold; the four-ebit curve crosses it between $t=720$ and $864$, while five ebits remain below threshold.}
\end{figure}

\textit{Dynamical embezzlement.---} The mean charge probabilities define the charge-resolved flattened spectrum $\overline{\widetilde\rho}_A(t)=\bigoplus_q[\bar p_q(t)/d_A(q)]\mathbb I_{d_A(q)}$. We evaluate the same exact optimization in Eq.~\eqref{eq:finite-fidelity-optimum} directly from the charge-sector weights and degeneracies, without forming the exponentially large matrix. Because Eq.~\eqref{eq:task} changes only at integer target dimensions, we denote by $\widetilde E_{\rm borrow}(\epsilon)$ its monotone interpolation in $\log_2m$. Figure~\ref{fig:operational}(a) shows that $\widetilde E_{\rm borrow}(0.5)$ follows the same fourth-root scale as $s_Q(t,L)$ and grows past four ebits for $L=144$.

The fixed-target conversion in Fig.~\ref{fig:operational}(b) gives a discrete threshold test. At $L=144$, the squared fidelity for borrowing four ebits crosses the required $0.75$ threshold between $t=720$ and $864$. Across the plotted temporal range, the normalized two-copy upper bound on unresolved within-sector nonuniformity falls from one to the few-percent scale~\cite{Note1}. Charge spreading creates the broad spectral envelope, and within-sector scrambling makes that envelope an increasingly effective catalyst.

To check that the average charge envelope also describes finite-size circuit trajectories, we employ direct wave-function simulations using TenCirPauli and TensorCircuit-NG which reproduce the charge-resolved behavior~\cite{zhang2023tensorcircuit,zhang2026tensorcircuitng}. 

This structure suggests a measurement protocol for a quantum simulator. Repeated computational-basis measurements of subsystem $A$ give the charge histogram $\bar p_q(t)$ and hence the broad envelope $\overline{\widetilde\rho}_A(t)$. Sector-resolved randomized measurements, or a swap measurement on two prepared copies, give the block purities that bound the remaining fine structure. Classical majorization post-processing then converts these measured quantities into the target fidelity for each $m$, producing the threshold curve in Fig.~\ref{fig:operational}(b) without full entanglement-spectrum tomography.

Therefore, this chronology separates two capabilities of the same entangled state. Ballistic dynamics establishes the leading volume-law entropy on the $O(L)$ timescale, whereas Figs.~\ref{fig:hydrodynamics} and~\ref{fig:operational} probe later diffusive transport toward the $O(\sqrt L)$ borrowing law at $t=O(L^2)$. The entropy records the leading entanglement volume, whereas the slow redistribution of the spectrum fixes how large a target can be lent and returned at fixed disturbance, turning hydrodynamics into a concrete quantum-information task.

\textit{Conclusions and discussion.---} Conserved-charge transport activates entanglement embezzlement by redistributing probability in the logarithmic Schmidt-rank coordinate. Away from half filling, charge bias converts $O(\sqrt L)$ number fluctuations into an $O(\sqrt L)$ finite-error borrowing resource, while particle--hole symmetry removes this growing contribution. During random-circuit evolution, diffusion builds the corresponding width as $t^{1/4}$. This separation connects transport universality to operational resource activation. Here conservation constrains the state dynamics, while extraction permits unrestricted LOCC; charge-respecting LOCC extraction for entanglement embezzlement remains a future problem.

\begin{acknowledgments}
\textbf{Acknowledgments.---}GPT-6 and GPT-5.6 assisted with parts of the code development and analytical derivations. All results were verified by the authors. SXZ was supported by the National Natural Science Foundation of China (No. 12574546), Quantum Science and Technology-National Science and Technology Major Project (No. 2024ZD0301700), and the Chinese Academy of Sciences (No. XDB1680201 and No. YSBR-150). SL was supported by the Gordon and Betty Moore Foundation (grant GBMF8685) towards the Princeton theory program, the Gordon and Betty Moore Foundation’s EPiQS Initiative (grant GBMF11070), the Global Collaborative Network Grant at Princeton University, the Simons Investigator Grant No. 404513, the NSF MRSEC (grant DMR-2011750), the Simons Collaboration on New Frontiers in Superconductivity (grant SFI-MPS-NFS-00006741-01), the Princeton Catalysis Initiative (PCI), the Schmidt Transformative Technology Fund at Princeton University. YQC was supported by the National Natural Science Foundation of China (No. 12504599), Quantum Science and Technology-National Science and Technology Major Project (No. 2025ZD0300802), and Science Challenge Project (No. TZ2025017).
\end{acknowledgments}

\bibliography{ref}

\end{document}


\setcounter{secnumdepth}{2}

\title{Supplemental Material for ``Entanglement Embezzlement from Diffusive Hydrodynamics''}
\author{Shi-Xin Zhang}
\email{shixinzhang@iphy.ac.cn}
\affiliation{Institute of Physics, Chinese Academy of Sciences, Beijing 100190, China}
\author{Shuo Liu}
\email{sl6097@princeton.edu}
\affiliation{Department of Physics, Princeton University, Princeton, New Jersey 08544, USA}
\author{Yu-Qin Chen}
\email{yqchen@gscaep.ac.cn}
\affiliation{Graduate School of China Academy of Engineering Physics, Beijing 100193, China}
\date{September 21, 2026}
\maketitle

This Supplemental Material gives the analytical derivations and the numerical evidence supporting the Letter. Sections~I--III establish the finite-spectrum conversion theorem, the Gaussian distribution across logarithmic Schmidt-rank, and the fixed-charge equilibrium result. Section~IV gives an independent finite-size convergence benchmark for the exact fixed-charge mean marginal. Section~V states the conditional dynamical implication, Section~VI derives the charge-adapted two-replica calculation, and Section~VII analyzes the fixed-transport control experiment. Natural logarithms are used unless $\log_2$ is written explicitly. For asymptotic sequences, $a_L=O(g_L)$, $a_L=o(g_L)$, $a_L\sim g_L$, and $a_L=\Theta(g_L)$ mean, respectively, that $|a_L/g_L|$ is bounded, $a_L/g_L\to0$, $a_L/g_L\to1$, and $|a_L/g_L|$ is bounded above and below by positive constants.

\section{Exact catalytic conversion problem}

The asymptotic analysis rests on an exact finite-dimensional conversion problem. We begin by identifying the pure-output optimum and proving the concave-majorant construction used to compute it, before relating that quantity to the unrestricted mixed-output trace error.

Let $p=(p_1,\ldots,p_d)$ be the nonincreasing Schmidt-probability vector of a pure catalyst $|\psi\rangle_{A_{\rm c}B_{\rm c}}$ shared by Alice and Bob, with Schmidt rank $d$. They append unentangled target registers $A_{\rm t}$ and $B_{\rm t}$ and seek $|\psi\rangle_{A_{\rm c}B_{\rm c}}\otimes|\Phi_m\rangle_{A_{\rm t}B_{\rm t}}$, where $|\Phi_m\rangle=m^{-1/2}\sum_{j=1}^m|j\rangle_{A_{\rm t}}|j\rangle_{B_{\rm t}}$ is maximally entangled with Schmidt rank $m$. Alice's local laboratory is $A_{\rm c}A_{\rm t}$ and Bob's is $B_{\rm c}B_{\rm t}$: local operations and classical communication (LOCC) refer throughout to this Alice--Bob partition, not to the catalyst--target tensor-product decomposition. Each party may therefore couple the catalyst and target registers held on that side, and the parties may exchange classical messages. The target Schmidt vector $v$ is obtained by replacing every $p_i$ by $m$ copies of $p_i/m$. We regard both $p$ and $v$ as vectors of the common dimension $M=md$, padding $p$ with zeros. Pure-state transformations under unrestricted LOCC are characterized by majorization~\cite{nielsen1999conditions}: $r\succ p$ means $\sum_{i=1}^k r_i\geq\sum_{i=1}^k p_i$ for every $k<M$, with equality at $k=M$. Among pure states reachable from $p$, the optimal root fidelity to $v$ is~\cite{vidal2000approximate,zanoni2024complete}
\begin{equation}
 F_m^\star=\max_{r\succ p}\sum_i\sqrt{r_iv_i},
 \qquad P_m^\star=\sqrt{1-(F_m^\star)^2}.
 \label{eq:pureopt}
\end{equation}
For density matrices, we use root fidelity $F(\rho,\sigma)=\|\sqrt\rho\sqrt\sigma\|_1$, purified distance $P(\rho,\sigma)=\sqrt{1-F(\rho,\sigma)^2}$, and trace distance $T(\rho,\sigma)=\tfrac12\|\rho-\sigma\|_1$, where $\|X\|_1=\Tr\sqrt{X^\dagger X}$ is the trace norm. Equation~\eqref{eq:pureopt} is the corresponding classical root fidelity between Schmidt vectors.

Define cumulative masses $A_k=\sum_{i=1}^kp_i$ and $B_k=\sum_{i=1}^kv_i$. The least concave majorant of the polygon through $(B_k,A_k)$ has blocks with increments $(\Delta B_j,\Delta A_j)$. Pooling adjacent blocks whenever their slopes increase gives
\begin{equation}
 F_m^\star=\sum_j\sqrt{\Delta A_j\Delta B_j}.
 \label{eq:pav}
\end{equation}
The pooling step is the following. Initially, every interval $(B_{k-1},B_k)$ is one block, with slope $\alpha_k=p_k/v_k$. Scan neighboring blocks from left to right. Whenever two adjacent slopes increase, $\alpha_j<\alpha_{j+1}$, merge those blocks into one; its increments become $\Delta B_j+\Delta B_{j+1}$ and $\Delta A_j+\Delta A_{j+1}$, so its new slope is the corresponding weighted average. Repeat this merge-and-rescan operation until the block slopes are nonincreasing. The final blocks are consecutive groups of indices. Each group becomes one straight segment of the least-concave majorant, and the number of blocks is the number left after all merges.
We now verify reachability. For any probability vector $u$, write its root fidelity to the target as $F(u)=\sum_i\sqrt{u_iv_i}$. On block $j$, set $\alpha_j=\Delta A_j/\Delta B_j$ and $r_i=\alpha_jv_i$. Concavity of the majorant makes the slopes $\alpha_j$ nonincreasing. Since $v_i$ is nonincreasing, $r_i$ is also nonincreasing. Define $R_k=\sum_{i=1}^kr_i$. This cumulative mass is the value of the majorant at $B_k$, so $R_k\geq A_k$ for every $k$, with equality at each hull contact; hence $r\succ p$ and the candidate is reachable.

To prove optimality, note that the fidelity gradient at the candidate is $g_i=\frac12\sqrt{v_i/r_i}=1/(2\sqrt{\alpha_j})$ on block $j$. Thus $g_i$ is constant inside a block and nondecreasing between blocks. For any other feasible vector $s$, define $S_k=\sum_{i=1}^ks_i$ and $E_k=S_k-R_k$. Whenever $g_{k+1}-g_k>0$, the hull has a contact and therefore $R_k=A_k$, while feasibility gives $S_k\geq A_k$ and hence $E_k\geq0$. Since $E_M=0$ at the padded target dimension $M$, discrete summation by parts gives
\begin{equation}
 \sum_{i=1}^{M}g_i(s_i-r_i)
 =-\sum_{k=1}^{M-1}(g_{k+1}-g_k)E_k\leq0.
 \label{eq:kktproof}
\end{equation}
Concavity of the fidelity then implies $F(s)\leq F(r)+\bm g\cdot(s-r)\leq F(r)$. On block $j$, the optimal fidelity contribution is $\sum_{i\in j}\sqrt{\alpha_j}v_i=\sqrt{\Delta A_j\Delta B_j}$, proving Eq.~\eqref{eq:pav}. This is the block solution of Vidal, Jonathan, and Nielsen~\cite{vidal2000approximate}, written as a least-concave-majorant problem. The pooling procedure is the pool-adjacent-violators (PAV) algorithm~\cite{busing2022pava}. For the fixed-charge spectra, the calculation acts on distinct eigenvalue plateaus and their rank boundaries. This compressed representation avoids constructing the exponentially large list of Schmidt levels and makes the exact conversion curve accessible for the large Hilbert spaces used below.

Let $T_m^\star$ be the minimum trace distance to the pure catalyst-plus-target state among all, possibly mixed, reachable outputs. The pure output attaining Eq.~\eqref{eq:pureopt} gives $T_m^\star\leq P_m^\star$. Conversely, measuring the projector onto the target yields $T_m^\star\geq1-(F_m^\star)^2=(P_m^\star)^2$. Hence
\begin{equation}
 (P_m^\star)^2\leq T_m^\star\leq P_m^\star.
 \label{eq:tracebracket}
\end{equation}
For a fixed allowed error $0<\epsilon<1$, define the purified-error yield
\begin{equation}
 E_{\rm borrow}(\epsilon)\equiv B_\epsilon^P=\max\{\log_2m:P_m^\star\leq\epsilon\}.
 \label{eq:borrowdefinition}
\end{equation}
The notation $E_{\rm borrow}$ is used in the Letter; $B_\epsilon^P$ is retained below where it keeps the purified- and trace-error formulas compact. This yield is directly achievable at trace error $\epsilon$.

When a continuous curve is useful for displaying finite-size trends, $\widetilde B_\epsilon^P$ denotes the monotone interpolation of the integer-target values in $\log_2m$.

For analytic stability, let $C_p(r)$ be the piecewise-linear cumulative Schmidt probability. The majorization defect for borrowing an $m$-dimensional target is
\begin{equation}
 \delta_m(p)=\max_{r\geq1}\left[C_p(r)-C_p(r/m)\right].
 \label{eq:defect}
\end{equation}
For a given rank cutoff $r$, the difference $C_p(r)-C_p(r/m)$ is the excess probability carried by the largest $r$ source coefficients over the corresponding target coefficients. Its maximum $\delta_m$ is therefore the worst cumulative-rank shortfall created by borrowing the target: $\delta_m=0$ means that every majorization constraint is satisfied exactly, while a larger value quantifies a stronger spectral obstruction. Standard finite-error conversion bounds turn this spectral deficit into operational guarantees, $\delta_m^2/2\leq T_m^\star\leq\sqrt{2\delta_m}$. Moreover, writing $\delta_m(\rho)$ for the defect of the descending eigenvalue vector of a density matrix $\rho$, any two density matrices $\rho$ and $\sigma$ obey
\begin{equation}
 |\delta_m(\rho)-\delta_m(\sigma)|\leq\tfrac12\|\rho-\sigma\|_1.
 \label{eq:lipschitz}
\end{equation}
For a fixed rank $r$, the difference in Eq.~\eqref{eq:defect} is a linear spectral functional with coefficients in $[0,1]$. Its variation is bounded by half the spectral $\ell_1$ distance; eigenvalue variation then proves Eq.~\eqref{eq:lipschitz}.

The purified optimization is uniformly stable as well. Let $\rho$ and $\sigma$ have eigenvalues $(\lambda_i)$ and $(\mu_i)$, and choose aligned Schmidt-basis purifications $|\psi_\rho\rangle=\sum_i\sqrt{\lambda_i}|i\rangle|i\rangle$ and $|\psi_\sigma\rangle=\sum_i\sqrt{\mu_i}|i\rangle|i\rangle$ such that $P(|\psi_\rho\rangle,|\psi_\sigma\rangle)=P(\rho,\sigma)$. Fix $m$ and let $\Lambda_m^\sigma$ be an optimal protocol for $|\psi_\sigma\rangle$. Contractivity gives a distance at most $P(\rho,\sigma)$ between the outputs of $\Lambda_m^\sigma$ on the two input catalysts. The ideal outputs $|\psi_\rho\rangle\otimes|\Phi_m\rangle$ and $|\psi_\sigma\rangle\otimes|\Phi_m\rangle$ are also separated by $P(\rho,\sigma)$. Applying the triangle inequality to these two input/output changes gives $P_m^\star(\rho)\leq P_m^\star(\sigma)+2P(\rho,\sigma)$. Exchanging $\rho$ and $\sigma$ gives the reverse inequality, and hence
\begin{equation}
 \sup_m|P_m^\star(\rho)-P_m^\star(\sigma)|
 \leq2P(\rho,\sigma)\leq2\sqrt{2T(\rho,\sigma)}.
 \label{eq:purifiedcontinuity}
\end{equation}
For the final inequality, $P(\rho,\sigma)^2=(1-F)(1+F)\leq2(1-F)\leq2T(\rho,\sigma)$ by the Fuchs--van de Graaf inequality. Thus
\begin{equation}
 P(\rho,\sigma)\leq\sqrt{2T(\rho,\sigma)}.
 \label{eq:purifiedtracebound}
\end{equation}
The bound is independent of $m$, so if $T(\rho_n,\sigma_n)\to0$ then $\sup_m|P_m^\star(\rho_n)-P_m^\star(\sigma_n)|\to0$; this transfers spectral limits uniformly across all target dimensions.

\section{Gaussian logarithmic Schmidt-rank spectrum and the conversion curve}

\subsection{Continuum picture}

Sample a Schmidt index $I$ with probability $p_I$ and write $R=\ln I$. When the probability mass has an approximately Gaussian profile of center $c_n$ and width $s_n$ in $R$,
\begin{equation}
 C_n\!\left(\ee^{c_n+s_nz}\right)\simeq\Phi(z),
 \label{eq:continuumrankcdf}
\end{equation}
where $C_n(r)$ is the cumulative mass of the largest $r$ Schmidt coefficients. Tensoring an $m_n$-dimensional maximally entangled target, whose uniform Schmidt-probability vector is $u_{m_n}=(1/m_n,\ldots,1/m_n)$, gives the exact identity $C_{p\otimes u_{m_n}}(r)=C_p(r/m_n)$, so the target profile is translated to the right by $\ln m_n$ in $R$. If $\ln m_n/s_n\to x$, the optimal majorization rearrangement compares two Gaussian profiles separated by $x$, and their square-root (Hellinger) overlap is
\begin{equation}
 F_{m_n}^\star\simeq\int_{-\infty}^{\infty}\sqrt{\phi(z)\phi(z+x)}\,dz
 =\ee^{-x^2/8}.
 \label{eq:continuumfidelity}
\end{equation}
The limiting majorization defect is likewise the largest Gaussian mass in an interval of length $x$, namely $2\Phi(x/2)-1$. This is the logarithmic Schmidt-rank-coordinate picture summarized in Fig.~1(a) of the Letter.

\subsection{Rigorous logarithmic Schmidt-rank theorem}

A finite entanglement spectrum is discrete and may be highly block structured, so its cumulative rank function is a staircase. The theorem below shows that uniform convergence of this cumulative profile is sufficient; no smoothness of individual eigenvalues is required. We write $\Phi(z)=\Pr\{Z\leq z\}$ and $\phi(z)=(2\pi)^{-1/2}\ee^{-z^2/2}$ for the cumulative distribution and density of a standard normal variable.

\begin{theorem}[Gaussian catalytic conversion]
Let $p^{(n)}$ be finite nonincreasing probability vectors after zero entries are omitted, and let $C_n(r)$ be their linearly interpolated cumulative mass, extended by one beyond the rank. Suppose that real centers $c_n$ and scales $s_n\to\infty$ satisfy
\begin{equation}
 A_n(z):=C_n\!\left(\ee^{c_n+s_nz}\right)\longrightarrow\Phi(z)
 \label{eq:rankcdf}
\end{equation}
uniformly in $z$. For every sequence of integer targets with $\ln m_n/s_n\to x\in[0,\infty)$,
\begin{align}
 F_{m_n}^\star&\longrightarrow\ee^{-x^2/8},
 &P_{m_n}^\star&\longrightarrow\sqrt{1-\ee^{-x^2/4}},
 \label{eq:theoremcurve}\\
 \delta_{m_n}&\longrightarrow2\Phi(x/2)-1.
 \label{eq:theoremdefect}
\end{align}
\end{theorem}

\paragraph{Proof.} For the ideal target vector $v=p\otimes u_{m_n}$, each $p_i/m_n$ occurs $m_n$ times, hence $C_v(r)=C_n(r/m_n)$ exactly under linear interpolation. With $z=(\ln r-c_n)/s_n$ and $x_n=\ln m_n/s_n$, the majorization defect is the supremum of $A_n(z)-A_n(z-x_n)$. Uniform convergence in Eq.~\eqref{eq:rankcdf} gives Eq.~\eqref{eq:theoremdefect}, because the largest normal mass in an interval of length $x$ is $2\Phi(x/2)-1$.

For the fidelity, define the generalized inverse $C_n^{-1}(v)=\inf\{r:C_n(r)\geq v\}$ on the positive source support. The target Lorenz polygon against the source cumulative is
\begin{equation}
 H_n(v)=C_n\!\left(m_nC_n^{-1}(v)\right),
 \qquad 0\leq v\leq1.
 \label{eq:finiteLorenz}
\end{equation}
For every $0<v<1$, Eq.~\eqref{eq:rankcdf} and inverse-distribution convergence give $[\ln C_n^{-1}(v)-c_n]/s_n\to\Phi^{-1}(v)$. Multiplication of rank by $m_n$ adds $x_n$ to this normalized coordinate. A second application of Eq.~\eqref{eq:rankcdf} gives pointwise convergence in the open interval, while both finite and limiting functions have endpoints zero and one. Since $H_n$ is nondecreasing and the limiting function is continuous on the closed interval, a finite-grid monotonicity argument upgrades the convergence to uniform convergence:
\begin{equation}
 H_n(v)\longrightarrow H_x(v)=\Phi\!\left[\Phi^{-1}(v)+x\right].
 \label{eq:lorenzlimit}
\end{equation}
It remains to pass this limit through the concave-envelope optimization. Let $G_n$ be the least concave majorant of $H_n$. With $z=\Phi^{-1}(v)$ and $\phi$ the standard normal density, $H_x'(v)=\phi(z+x)/\phi(z)=\ee^{-xz-x^2/2}$ is nonincreasing for $x\geq0$, so $H_x$ is concave. If $\|H_n-H_x\|_\infty\leq e_n$, then $H_x+e_n$ is a concave majorant of $H_n$, whereas $G_n\geq H_n\geq H_x-e_n$. It follows that $\|G_n-H_x\|_\infty\leq e_n\to0$.

At an interior point, the derivative of a concave function lies between its backward and forward secant slopes. Taking $n\to\infty$ at fixed secant width and then shrinking that width proves $G_n'\to H_x'$ almost everywhere. Both derivatives are nonnegative and integrate to one. Scheff\'e's lemma therefore gives $\|G_n'-H_x'\|_1\to0$ without losing slope mass at the endpoints. Finally,
\begin{equation}
 \int_0^1|\sqrt{G_n'}-\sqrt{H_x'}|\,dv
 \leq\left(\int_0^1|G_n'-H_x'|\,dv\right)^{1/2}\longrightarrow0.
 \label{eq:sqrtderivative}
\end{equation}
Equation~\eqref{eq:pav} therefore becomes
\begin{align}
 \lim_{n\to\infty}F_{m_n}^\star
 &=\int_0^1\sqrt{H_x'(u)}\,du \notag\\
 &=\int_{-\infty}^{\infty}\sqrt{\phi(z)\phi(z+x)}\,dz
 =\ee^{-x^2/8}.
 \label{eq:hellinger}
\end{align}
Equation~\eqref{eq:hellinger} proves the fidelity limit and hence
\begin{equation}
 P_{m_n}^\star\to P_{\rm G}(x)=\sqrt{1-\ee^{-x^2/4}},
 \qquad B_\epsilon^P=\frac{2s_n}{\ln2}\sqrt{-\ln(1-\epsilon^2)}+o(s_n).
 \label{eq:gausslaw}
\end{equation}
To justify the inversion, fix $0<\epsilon<1$ and write $x_\epsilon=2\sqrt{-\ln(1-\epsilon^2)}$. For arbitrary $\eta>0$, choose integer targets nearest to $\exp[(x_\epsilon\pm\eta)s_n]$. The limiting curve is strictly increasing, so monotonicity in $m$ places the finite threshold between these targets for all sufficiently large $n$. Sending $\eta\to0$ proves the yield formula. Restricting $m$ to powers of two changes the yield by less than one Bell pair and leaves the leading coefficient unchanged. If $B_\epsilon^T=\max\{\log_2m:T_m^\star\leq\epsilon\}$ denotes the unrestricted trace-error yield, Eq.~\eqref{eq:tracebracket} gives
\begin{equation}
 B_\epsilon^P\leq B_\epsilon^T\leq B_{\sqrt\epsilon}^P.
 \label{eq:yieldbracket}
\end{equation}

This proves that a Gaussian probability profile in the logarithmic Schmidt-rank coordinate fixes the full leading purified-error curve and its exact fixed-error coefficient. The result is insensitive to charge-sector staircases and to the smoothness of individual eigenvalues.

\subsection{A sufficient condition for the logarithmic Schmidt-rank coordinate}

The fixed-charge physics most directly controls the information content $K_n=-\ln p_I^{(n)}$, where $I$ is sampled with probability $p_I^{(n)}$. Suppose that for the same centers and a diverging scale,
\begin{equation}
 \frac{K_n-c_n}{s_n}\Rightarrow Z,
 \qquad Z\sim\mathcal N(0,1).
 \label{eq:clt}
\end{equation}
This information central limit theorem (CLT) implies the logarithmic Schmidt-rank hypothesis in Eq.~\eqref{eq:rankcdf}. Indeed, for any real $a$ and $u>0$, counting eigenvalues above and below the threshold $\ee^{-a}$ gives
\begin{equation}
 \Pr\{K_n\leq a\}\leq C_n(\ee^a)
 \leq\Pr\{K_n\leq a+u\}+\ee^{-u}.
 \label{eq:counting}
\end{equation}
The first inequality follows because every eigenvalue at least $\ee^{-a}$ lies among the first $\ee^a$ ranks; eigenvalues smaller than $\ee^{-a-u}$ contribute at most $\ee^a\ee^{-a-u}=\ee^{-u}$ to those ranks, proving the second. Choose $u_n\to\infty$ with $u_n/s_n\to0$ and put $a=c_n+s_nz$. Since the Gaussian cumulative is continuous, weak convergence is uniform in $z$, and Eq.~\eqref{eq:counting} yields Eq.~\eqref{eq:rankcdf}.

Up to interpolation at noninteger rank, $A_n$ is the cumulative distribution of the probability-weighted logarithmic Schmidt-rank coordinate $R_n=\ln I$. Thus the two coordinates have the same centered limit on scale $s_n$, although this does not assert pointwise equality of $K_n$ and $R_n$. Nor does it determine the coefficient of a fixed-width octave: the $\Theta(s_n^{-1})$ law below additionally uses a local CLT and the fixed-charge block geometry. If $s_n$ remains bounded, as at half filling, this equivalence is unavailable and the separate rank argument below applies.

\section{Fixed-charge Haar states}

We now specialize the general theorem to fixed-charge equilibrium. At an unequal cut, a typical fixed-charge Haar marginal approaches an exactly known mean marginal; the mean spectrum in turn obeys an information CLT. Uniform operational continuity links these two facts and transfers the complete conversion law to typical pure states. Half filling requires a separate argument because its thermodynamic force vanishes.

Consider $L$ qubits with occupation operators $n_i=|1\rangle_i\langle1|$ and conserved total charge $Q=\sum_{i=1}^L n_i=N$. Divide the chain into a subsystem $A$ of $L_A=fL$ qubits and its complement $B$ of $L_B=L-L_A$ qubits, with fixed $0<f<1/2$. The subsystem charge is $Q_A=\sum_{i\in A}n_i$, and its eigenvalue is denoted by $q$. At fixed filling $\nu=N/L$, write
\begin{equation}
 a_q=\binom{L_A}{q},\qquad b_q=\binom{L_B}{N-q},\qquad \mathcal D=\binom{L}{N}.
\end{equation}
The mean marginal of a Haar-random vector in the fixed-charge subspace is
\begin{equation}
 \omega_A=\bigoplus_q\frac{b_q}{\mathcal D}\mathbb I_{a_q},
 \qquad \bar p_q=\frac{a_qb_q}{\mathcal D}.
 \label{eq:meanmarginal}
\end{equation}
Here $a_q$ and $b_q$ are the dimensions of the charge-$q$ subsystem sector and the compatible charge-$(N-q)$ environment sector, respectively, $\mathcal D$ is the fixed-charge Hilbert-space dimension, $\mathbb I_{a_q}$ is the identity on the subsystem sector, and $\bar p_q$ is the probability of observing $Q_A=q$.
\subsection{Concentration about the mean marginal}

This subsection establishes why the exactly calculable mean marginal is representative of a typical fixed-charge pure state. The mean marginal is block flat in each subsystem-charge sector, whereas an individual Haar-random state has fluctuating sector weights and nonuniform matrices within the blocks. Dirichlet concentration controls the former fluctuations and induced-Wishart estimates control the latter, showing that the full reduced state approaches the mean marginal exponentially closely at a fixed unequal cut. Operational continuity then transfers the conversion curves obtained from the mean spectrum to typical pure states, so the remaining equilibrium analysis can be carried out on the simpler block-flat matrix.

Sampling uniformly from the fixed-charge subspace means sampling with respect to its unitarily invariant Haar measure. Let $\rho_A$ be the reduced density matrix of such a pure state. Its sector weights $w_q$ have a Dirichlet distribution with parameters $a_qb_q$, whose density on $\sum_qw_q=1$ is proportional to $\prod_qw_q^{a_qb_q-1}$. Conditional on $w_q$, the normalized block $\sigma_q$ is an induced random state obtained by tracing the $b_q$-dimensional factor from a Haar-random vector on $\mathbb C^{a_q}\otimes\mathbb C^{b_q}$. Thus $\rho_A=\bigoplus_qw_q\sigma_q$, and a blockwise triangle inequality gives
\begin{equation}
 T(\rho_A,\omega_A)
 \leq\frac12\sum_q|w_q-\bar p_q|
 +\sum_qw_qT\!\left(\sigma_q,\frac{\mathbb I_{a_q}}{a_q}\right).
 \label{eq:blocktriangle}
\end{equation}
Let $n_{\rm sec}$ be the number of allowed subsystem-charge sectors, so $n_{\rm sec}\leq L_A+1$. For the Dirichlet distribution, $\Var w_q=\bar p_q(1-\bar p_q)/(\mathcal D+1)$. Cauchy--Schwarz therefore implies
\begin{equation}
 \mathbb E\sum_q|w_q-\bar p_q|
 \leq\sqrt{n_{\rm sec}\sum_q\Var w_q}
 \leq\sqrt{\frac{n_{\rm sec}}{\mathcal D+1}}.
 \label{eq:dirichlet}
\end{equation}
The normalized direction in each sector is independent of its Dirichlet weight, and its expected purity is $(a_q+b_q)/(a_qb_q+1)$. With the Hilbert--Schmidt norm $\|X\|_2=[\Tr(X^\dagger X)]^{1/2}$, the inequality $\|X\|_1\leq\sqrt{a_q}\|X\|_2$ and Jensen's inequality give
\begin{equation}
 \mathbb E\,T\!\left(\sigma_q,\frac{\mathbb I_{a_q}}{a_q}\right)
 \leq\frac12\sqrt{a_q\mathbb E\Tr\sigma_q^2-1}
 =\frac12\sqrt{\frac{a_q^2-1}{a_qb_q+1}}.
 \label{eq:wishart}
\end{equation}
Combining Eqs.~\eqref{eq:blocktriangle}--\eqref{eq:wishart}, and capping a block trace distance by one, gives
\begin{equation}
 \mathbb E\,T(\rho_A,\omega_A)\leq\frac12\sqrt{\frac{n_{\rm sec}}{\mathcal D+1}}+\frac12\sum_q\bar p_q\min\!\left\{2,\sqrt{\frac{a_q^2-1}{a_qb_q+1}}\right\},
 \label{eq:typicality}
\end{equation}
where the minimum uses the elementary upper bound $T\leq1$ whenever the dimension-ratio estimate is weaker.

We next show explicitly why Eq.~\eqref{eq:typicality} vanishes. Fix $f<1/2$ and let the filling stay in a compact subset of $(0,1)$. Choose a sufficiently small $\delta>0$ and define $\mathcal W_L=\{|q-\nu L_A|\leq\delta L\}$. With the binary entropy $h(x)=-x\ln x-(1-x)\ln(1-x)$, uniform Stirling estimates give
\begin{equation}
 \frac1L\ln\frac{b_q}{a_q}
 =(1-f)h\!\left(\frac{\nu L-q}{(1-f)L}\right)
 -fh\!\left(\frac{q}{fL}\right)+o(1).
 \label{eq:dimensionrate}
\end{equation}
At the mean charge, the right-hand side is $(1-2f)h(\nu)>0$. Continuity therefore permits a choice of $\delta$ and $c_1>0$ such that $b_q/a_q\geq\ee^{c_1L}$ uniformly in $\mathcal W_L$. Hoeffding's inequality for sampling $L_A=fL$ sites without replacement gives the explicit hypergeometric tail bound $\sum_{q\notin\mathcal W_L}\bar p_q\leq2\exp(-2\delta^2L/f)$. On $\mathcal W_L$, the square root in Eq.~\eqref{eq:typicality} is at most $\sqrt{a_q/b_q}\leq\ee^{-c_1L/2}$; outside it, the cap bounds the contribution by twice the tail weight. The Dirichlet term is exponentially small because $\mathcal D=\exp[Lh(\nu)+o(L)]$. Hence constants $C,c>0$ exist such that
\begin{equation}
 \mathbb E\,T(\rho_A,\omega_A)\leq C\ee^{-cL}.
 \label{eq:typicalityexponential}
\end{equation}
Markov's inequality proves $T(\rho_A,\omega_A)\to0$ in probability. Equation~\eqref{eq:purifiedcontinuity} then transfers the entire purified-error curve uniformly in $m$ from the mean marginal to a typical Haar marginal. Accordingly, a conversion evaluated on $\omega_A$ refers to a pure catalyst whose Schmidt-probability vector is the eigenvalue spectrum of $\omega_A$.

\begin{figure}[t]
 \centering
 \includegraphics[width=0.52\textwidth]{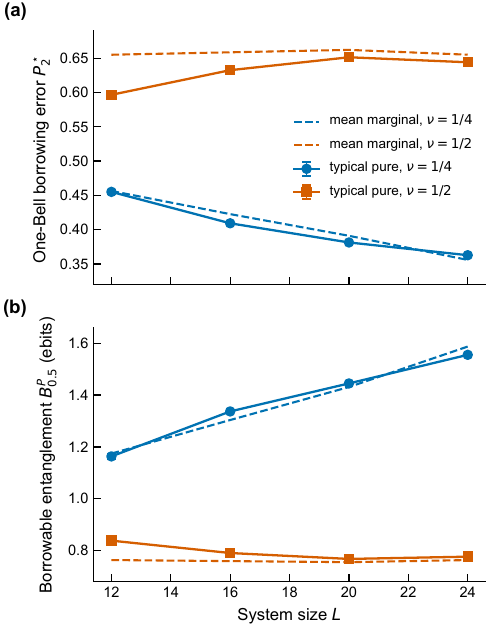}
 \caption{\label{fig:typicalSM}Full-state check of the fixed-charge typicality theorem. Markers and error bars are the means and standard errors of observables computed separately for 24 independent fixed-charge Haar pure states with $L_A=\lfloor L/3\rfloor$; the averaging is performed after computing each state's reduced-spectrum observable. These are the typical-state, or average-after-evaluation, quantities. Dashed curves are the corresponding mean-marginal, or evaluate-after-averaging, observables computed after first forming the exact ensemble-averaged marginal $\omega_A$. (a) The purified error $P_2^\star$ for borrowing one Bell pair decreases with size at $\nu=1/4$, while the half-filled value remains finite. (b) At global purified error $0.5$, the continuously interpolated borrowing yield $\widetilde B_{0.5}^P$ grows at $\nu=1/4$ and remains bounded at $\nu=1/2$. The comparison tests the finite-size agreement between typical-state and mean-marginal observables.}
\end{figure}

For an observable $O$ of the reduced-spectrum state, the two averaging orders used below are
\begin{equation}
 O_{\rm avg\,after\,eval}=\frac1{N_s}\sum_{s=1}^{N_s}O\!\left(\rho_A^{(s)}\right),
 \qquad
 O_{\rm eval\,after\,avg}=O(\omega_A),
 \qquad
 \omega_A=\frac1{N_s}\sum_{s=1}^{N_s}\rho_A^{(s)}.
 \label{eq:averagingorders}
\end{equation}
The first quantity evaluates $P_m^\star$ or the yield for each pure-state sample and then averages the resulting numbers. The second forms the mean marginal first and evaluates the same observable afterward. The markers in Fig.~\ref{fig:typicalSM} use the first order, and the dashed curves use the second order.

Figure~\ref{fig:typicalSM} thus provides a finite-size check of this mean-to-typical transfer. For the markers, $P_2^\star$ and $\widetilde B_{0.5}^P$ are evaluated for each sampled pure-state spectrum and then averaged; for the dashed curves, the state is averaged first to form $\omega_A$ and the observable is evaluated afterward. Their agreement is the finite-size evidence that the two operations become close in the unequal-cut typicality regime.

\subsection{Information central limit and equilibrium yield}

This subsection identifies the mechanism that turns charge fluctuations into a usable embezzlement yield. The sector masses of the mean marginal form a hypergeometric charge distribution, which is asymptotically Gaussian with width $\sigma_Q$. The relative-information variable measures the difference between the entanglement energy $K=-\ln\lambda$ and its linear charge contribution $\mu Q_A$; the tail bounds below show that this residual is negligible on the scale $|\mu|\sigma_Q$. Consequently, the entanglement energies and the logarithmic Schmidt-rank coordinate have the same Gaussian fluctuations, and the general conversion theorem immediately gives the full fidelity curve and the $O(\sqrt L)$ equilibrium yield.

We now prove the information CLT from relative-information tail bounds. Let $\tau_\nu=\operatorname{diag}(1-\nu,\nu)$ and $\mu=\ln[(1-\nu)/\nu]$, so $\tau_\nu^{\otimes L_A}=\ee^{-\mu Q_A}/Z_A$, where $Z_A=(1-\nu)^{-L_A}$ normalizes the product state. For density matrices $\rho$ and $\tau$, their quantum relative entropy is $D(\rho\Vert\tau)=\Tr[\rho(\ln\rho-\ln\tau)]$. For any commuting pair $\rho_A$ and $\tau_\nu^{\otimes L_A}$, let $\lambda$ be an eigenvalue of $\rho_A$ sampled with probability $\lambda$ and define its information content by $K=-\ln\lambda$. We then define the relative-information random variable
\begin{equation}
 J=\ln\rho_A-\ln\tau_\nu^{\otimes L_A}
 =-K+\mu Q_A+\ln Z_A,
 \label{eq:relativeinformation}
\end{equation}
sampled with the eigenvalues of $\rho_A$. The following elementary bounds control both tails:
\begin{align}
 \Pr_{\rho_A}(J<-a)&\leq\ee^{-a}, ~~
 \mathbb E_{\rho_A}J_-\leq\frac1e,
 \label{eq:negativeJ}\\
 \Pr_{\rho_A}(J>a)&\leq\frac{D(\rho_A\Vert\tau_\nu^{\otimes L_A})+1/e}{a}.
 \label{eq:positiveJ}
\end{align}
Here $J_-=\max\{-J,0\}$ and $J_+=\max\{J,0\}$. For completeness, let $p_i$ and $t_i$ be the simultaneous eigenvalues of $\rho_A$ and $\tau_\nu^{\otimes L_A}$ and write $\ell_i=p_i/t_i<1$ on the negative part of $J$. On $J<-a$, $p_i=t_i\ell_i\leq\ee^{-a}t_i$, which proves the first bound. Moreover, $t_i\ell_i\ln(1/\ell_i)\leq t_i/e$, proving the second. Since $\mathbb E J=D(\rho_A\Vert\tau_\nu^{\otimes L_A})=\mathbb E J_+-\mathbb E J_-$, Markov's inequality gives the last bound.

Let $\Pi_N$ be the projector onto the total-charge-$N$ subspace. The product Bernoulli state $\tau_\nu^{\otimes L}$ conditioned on total charge $N=\nu L+O(1)$ is the maximally mixed fixed-charge state $\Omega_N=\Pi_N/\mathcal D$. Therefore
\begin{align}
 D(\Omega_N\Vert\tau_\nu^{\otimes L})
 &=-\ln\Pr\{X=N\},\qquad X\sim\operatorname{Binomial}(L,\nu), \notag\\
 &=\tfrac12\ln[2\pi L\nu(1-\nu)]+O(1).
 \label{eq:canonicalrelativeentropy}
\end{align}
Data processing inequality under the partial trace gives $D(\omega_A\Vert\tau_\nu^{\otimes L_A})=O(\ln L)$. For fixed $\nu\ne1/2$, set
\begin{equation}
 s_L=|\mu|\sigma_Q,
 \qquad
 \sigma_Q^2=f(1-f)\nu(1-\nu)L+O(1).
 \label{eq:chargewidth}
\end{equation}
Since $s_L=\Theta(\sqrt L)$, Eqs.~\eqref{eq:negativeJ} and \eqref{eq:positiveJ} imply $J/s_L\to0$ in probability. With $\langle Q_A\rangle=\Tr(\rho_AQ_A)$ and $\operatorname{sgn}(\mu)=\mu/|\mu|$, Eq.~\eqref{eq:relativeinformation} gives
\begin{equation}
 \frac{K-c_L}{s_L}
 =\operatorname{sgn}(\mu)\frac{Q_A-\langle Q_A\rangle}{\sigma_Q}-\frac{J}{s_L},
 \qquad c_L=\mu\langle Q_A\rangle+\ln Z_A.
 \label{eq:informationdecomposition}
\end{equation}
The charge law $\bar p_q$ is hypergeometric, so its ordinary central limit theorem and Slutsky's theorem prove Eq.~\eqref{eq:clt} with scale $s_L$.

The force $\mu$ can also be read directly from adjacent eigenvalues. With $\lambda_q=b_q/\mathcal D$ and $K_q=-\ln\lambda_q$,
\begin{equation}
 K_{q+1}-K_q
 =\ln\frac{L_B-N+q+1}{N-q}
 =\mu+O\!\left(\frac{1+|q-\nu L_A|}{L}\right).
 \label{eq:rankforce}
\end{equation}
Let $\bar q\equiv\langle Q_A\rangle=\nu L_A$ denote the mean subsystem charge. Across a typical $O(\sqrt L)$ charge window, this yields $K_q=K_0+\mu(q-\bar q)+O(1)$, where $K_0$ is independent of $q$. The corresponding rank coordinate can be compared directly. For fixed $\nu<1/2$, let $r_q$ be the cumulative rank through the charge-$q$ block in decreasing-eigenvalue order. Since $a_{q+1}/a_q=\ee^{\mu}[1+O(L^{-1/2})]$, the last term dominates the cumulative rank and $R_q\equiv\ln r_q=\ln a_q+O(1)$. Meanwhile $K_q=\ln a_q-\ln\bar p_q$, and the hypergeometric local CLT gives $-\ln(\bar p_q/\bar p_{\bar q})=O(1)$ uniformly when $|q-\bar q|=O(\sqrt L)$. After choosing separate centers $K_0$ and $R_0$, therefore,
\begin{equation}
 (K_q-K_0)-(R_q-R_0)=O(1)=o(s_L).
 \label{eq:fixedchargerankequivalence}
\end{equation}
Particle--hole conjugation gives the same statement for fixed $\nu>1/2$. Thus $s_L=|\mu|\sigma_Q$ is the width of the probability envelope in the logarithmic Schmidt-rank coordinate under the fixed non-half-filling limit. Equation~\eqref{eq:rankforce} explains the mechanism, while Eqs.~\eqref{eq:canonicalrelativeentropy}--\eqref{eq:fixedchargerankequivalence} connect the information-content proof to the logarithmic Schmidt-rank-coordinate picture in Fig.~1(a) of the Letter.

Applying Theorem~1 and the typicality result gives, for every fixed $0<\epsilon<1$,
\begin{equation}
 B_\epsilon^P=
 \frac{2\sqrt{-\ln(1-\epsilon^2)}}{\ln2}
 \left|\ln\frac{1-\nu}{\nu}\right|
 \sqrt{f(1-f)\nu(1-\nu)L}+o(\sqrt L),
 \label{eq:equilibriumyield}
\end{equation}
both for the mean spectrum and in probability for typical fixed-charge Haar pure states. Equation~\eqref{eq:yieldbracket} gives the corresponding upper and lower leading bounds for the unrestricted trace-error yield.

\subsection{Dominant-scale weight}

This subsection translates the broad-spectrum picture into the fixed-width octave statistic used to diagnose universal embezzlement. Away from half filling, changing the rank by a factor of two crosses only finitely many neighboring charge blocks, and each central block carries probability of order $L^{-1/2}$. This gives $\eta^\star=\Theta(L^{-1/2})$. At half filling, neighboring blocks become nearly equal across a $\sqrt L$-wide central charge window, so one octave captures a finite amount of probability instead. The argument therefore connects the charge-sector geometry directly to the established rank-octave criterion and then transfers the result to typical pure states.

The factor-two rank-window statistic used in the Letter is
\begin{equation}
 \eta^\star(\rho_A)=\max_{\ell\geq0}\sum_{i=\ell}^{2\ell}p_i,
 \label{eq:etasupp}
\end{equation}
where $p_0\geq p_1\geq\cdots$ are the Schmidt probabilities, the indices are zero based, and the upper endpoint is clipped at the rank. Equivalently, if $C(r)$ linearly interpolates the cumulative probability of the first $r$ levels, then $\eta^\star=\max_{\ell\geq0}[C(2\ell+1)-C(\ell)]$. This is the established octave concentration controlling universal LOCC embezzlement~\cite{sierant2026nonlocalmagic}; here we derive how fixed charge sets its scale.

For fixed $\nu<1/2$, the mean-marginal eigenvalues $\lambda_q=b_q/\mathcal D$ decrease with $q$ throughout the hypergeometric typical window. Adjacent charge blocks obey
\begin{align}
 \frac{\lambda_{q+1}}{\lambda_q}&=\frac{N-q}{L_B-N+q+1}=\ee^{-\mu}[1+O(L^{-1/2})],\notag\\
 \frac{a_{q+1}}{a_q}&=\frac{L_A-q}{q+1}=\ee^{\mu}[1+O(L^{-1/2})],
 \label{eq:adjacentblocks}
\end{align}
uniformly for $|q-\nu L_A|=O(\sqrt L)$. Consequently, multiplying rank by two crosses only a filling-dependent $O(1)$ number of charge blocks. The hypergeometric local central limit theorem gives $\max_q\bar p_q=\Theta(L^{-1/2})$ and assigns $O(L^{-1/2})$ weight to every block within a fixed distance of its mean. Every factor-two rank interval therefore carries at most $O(L^{-1/2})$ central weight, while an interval placed inside a central block captures a filling-dependent positive fraction of that block. Hypergeometric tails are exponentially small outside an extensive typical window. Hence
\begin{equation}
 \eta^\star(\omega_A)=\Theta(L^{-1/2}),\qquad \nu\ne\frac12.
 \label{eq:etaoffhalf}
\end{equation}
Particle--hole conjugation gives the same result for $\nu>1/2$. The eigenvalue variation bound $\sum_i|\lambda_i(\rho_A)-\lambda_i(\omega_A)|\leq\|\rho_A-\omega_A\|_1$, together with Eq.~\eqref{eq:typicalityexponential}, transfers Eq.~\eqref{eq:etaoffhalf} to typical fixed-charge Haar states in probability.

At half filling, Eq.~\eqref{eq:adjacentblocks} approaches unity across the $O(\sqrt L)$ central charge window. A fixed factor-two change in rank then spans $\Theta(\sqrt L)$ adjacent central blocks. Uniform Stirling estimates map such a window to a fixed interval of the Gaussian charge coordinate and therefore give a strictly positive limiting lower bound on its probability. Since $\eta^\star\leq1$, this proves
\begin{equation}
 \eta^\star(\omega_A)=\Theta(1),\qquad \nu=\frac12,
 \label{eq:etahalf}
\end{equation}
and exponential typicality again transfers the result to typical pure states. Equations~\eqref{eq:etaoffhalf} and \eqref{eq:etahalf} prove the scaling contrast shown in Fig.~1(a) of the Letter.

\subsection{Half filling case}

This subsection treats the singular point where the charge fluctuations remain large but their linear spectral response disappears. At half filling the effective bias $\mu$ is zero, so the $O(\sqrt L)$ charge window does not produce an extensive width in the logarithmic Schmidt-rank coordinate. A rank bound and the Ky Fan principle then show directly that the finite-error borrowing yield stays bounded as $L$ grows.

At $\nu=1/2$, the charge variance is still extensive but the force $\mu$ vanishes. This case separates the existence of charge fluctuations from their ability to broaden the logarithmic Schmidt-rank profile. The linear term in Eq.~\eqref{eq:rankforce} vanishes. To quantify the remaining variation, write $q=L_A/2+\delta$ and $M=L_B/2$, so that $\delta=O(\sqrt L)$ across the typical charge window and $M=O(L)$. Equation~\eqref{eq:rankforce} becomes
\begin{equation}
 K_{q+1}-K_q=\ln\frac{M+\delta+1}{M-\delta}
 =\frac{2\delta+1}{M}+O\!\left(\frac{1+\delta^2}{M^2}\right).
 \label{eq:halfrankincrement}
\end{equation}
Summing from the central block to $\delta=O(\sqrt L)$ gives
\begin{equation}
 K_q-K_{L_A/2}=\frac{\delta^2}{M}+O\!\left(\frac{|\delta|+|\delta|^3}{M^2}\right)=O(1).
 \label{eq:halfrankvariation}
\end{equation}
The same estimate holds on the lower-charge side, with parity shifts changing only the $O(1)$ reference term. The quadratic variation across the typical charge window therefore remains $O(1)$. Boundedness at every fixed error below one follows from a rank bound. The mean marginal has rank $d=2^{L_A}$ and
\begin{equation}
 \chi_L=d\lambda_{\max}(\omega_A)
 =\frac{2^{L_A}\binom{L_B}{\lfloor L_B/2\rfloor}}{\binom{L}{L/2}}
 \longrightarrow(1-f)^{-1/2}.
 \label{eq:halfmax}
\end{equation}
The limit follows from the central-binomial Stirling formula, with floor functions handling parity-compatible sequences. Every reachable pure output has rank at most $d$. By the Ky Fan variational principle, its squared overlap with the target is bounded by the target's first-$d$ Schmidt mass, which is at most $d\lambda_{\max}/m=\chi_L/m$. Consequently,
\begin{equation}
 B_\epsilon^P\leq\log_2\frac{\chi_L}{1-\epsilon^2},
 \qquad B_\epsilon^T\leq\log_2\frac{\chi_L}{1-\epsilon}.
 \label{eq:halfyield}
\end{equation}
To make the typical-state transfer explicit, fix $\epsilon<1$. For purified error, choose an integer target dimension $M>(1-f)^{-1/2}/(1-\epsilon^2)$ with a strict margin; for trace error, use $M>(1-f)^{-1/2}/(1-\epsilon)$. The corresponding mean-spectrum error at $M$ then remains above the threshold by a positive margin. Equations~\eqref{eq:typicalityexponential} and \eqref{eq:purifiedcontinuity}, followed by monotonicity in $m$, show that typical Haar states have purified and trace yields bounded by $\log_2 M+o(1)$ at every fixed error below one.



\section{Equilibrium finite-size convergence}

The asymptotic Gaussian law describes the growing-scale limit, while the finite systems used in the numerical simulation retain discrete charge sectors and integer target dimensions. We therefore quantify how the finite-size ensemble-averaged mean marginal approaches the Gaussian conversion law and identify the separate effects of curve discretization and finite-size spectral corrections.

For each finite $L$, we evaluate the ensemble-averaged mean marginal $\omega_A$ in Eq.~\eqref{eq:meanmarginal} from its exact fixed-charge sector weights and degeneracies. Haar averaging makes this mean marginal block flat within each subsystem-charge sector. The spectrum being evaluated is the block-flat mean spectrum.

Define the scaled target coordinate and the target grid by
\begin{equation}
 x=\frac{\ln m}{s_L},\qquad s_L=|\mu|\sigma_Q,\qquad
 \mathcal G=\{0.25,0.5,1,1.5,2,3\}.
 \label{eq:tabletargetgrid}
\end{equation}
For each requested $x_{\rm req}\in\mathcal G$, the calculation sets $m=\max\{1,\operatorname{round}[\exp(x_{\rm req}s_L)]\}$, evaluates the actual coordinate $x_{\rm act}=\ln m/s_L$, and uses this same $x_{\rm act}$ in the exact mean-marginal calculation and in the Gaussian predictions. The Gaussian predictions are $P_{\rm G}(x)=\sqrt{1-\exp(-x^2/4)}$ and $\delta_{\rm G}(x)=2\Phi(x/2)-1$.

Table~\ref{tab:gaussiancheck} contains two exact--Gaussian curve discrepancies and one yield ratio. The third-column quantity is
\begin{equation}
 \Delta_P=\max_{x_{\rm req}\in\mathcal G}\left|P_{m(x_{\rm req})}^\star(\omega_A)-P_{\rm G}(x_{\rm act})\right|,
 \label{eq:tablePdifference}
\end{equation}
and the fourth-column quantity is
\begin{equation}
 \Delta_\delta=\max_{x_{\rm req}\in\mathcal G}\left|\delta_{m(x_{\rm req})}(\omega_A)-\delta_{\rm G}(x_{\rm act})\right|.
 \label{eq:tabledeltadifference}
\end{equation}
The final column gives $B_{0.5}^{P,{\rm disc}}/B_{0.5}^{P,{\rm lead}}$, where $B_{0.5}^{P,{\rm disc}}=\log_2m_\star$ is the yield obtained by optimizing over integer target dimensions $m$, and $B_{0.5}^{P,{\rm lead}}$ is the leading continuous Gaussian prediction. Thus the table compares the finite-size mean marginal with the Gaussian law at matched scaled targets, while the yield ratio records the effect of discrete target dimensions on the yield. These quantities decrease toward their asymptotic values with increasing size at all three fixed fillings.
\begin{table}[h]
\caption{\label{tab:gaussiancheck}Finite-size comparison of the ensemble-averaged mean marginal with the Gaussian conversion law for $L_A=L/3$. The mean marginal $\omega_A$ is evaluated exactly from the fixed-charge sector weights and degeneracies, and its spectrum is block flat within each charge sector. For each requested $x_{\rm req}\in\mathcal G$, the target dimension is the integer $m=\operatorname{round}[\exp(x_{\rm req}s_L)]$ and the actual coordinate is $x_{\rm act}=\ln m/s_L$. The third column gives $\Delta_P$, the maximum absolute difference between the purified error from this mean marginal and the Gaussian purified-error curve. The fourth column gives $\Delta_\delta$, the corresponding maximum absolute difference for the majorization defect. The final column gives the ratio of the exact discrete-target yield $B_{0.5}^{P,{\rm disc}}=\log_2m_\star$ to the leading continuous yield prediction at $\epsilon=0.5$.}
\begin{ruledtabular}
\begin{tabular}{ccccc}
$\nu$ & $L$ & $\Delta_P$ & $\Delta_\delta$ & $B_{0.5}^{P,{\rm disc}}/B_{0.5}^{P,\rm lead}$ \\
\hline
$1/4$ & 96   & 0.02809 & 0.02619 & $3/3.418$ \\
      & 6144 & 0.00252 & 0.00046 & $27/27.206$ \\
$1/6$ & 96   & 0.05102 & 0.03729 & $3/4.310$ \\
      & 6144 & 0.00665 & 0.00552 & $33/34.302$ \\
$1/12$& 96   & 0.09297 & 0.06813 & $4/4.762$ \\
      & 6144 & 0.01206 & 0.00976 & $37/37.902$ \\
\end{tabular}
\end{ruledtabular}
\end{table}

The calculation uses the fixed-charge sector weights and degeneracies in Eq.~\eqref{eq:meanmarginal}, represents them as integer multiplicity--eigenvalue blocks, and processes them with the compressed PAV construction. It evaluates an equilibrium mean-spectrum benchmark.

\section{Dynamical implication and stability interface}

Generic nonconserving chaotic circuits can create an intermediate universal-embezzling spectrum through a ballistically moving and broadening packet of weight in the logarithmic Schmidt-rank coordinate~\cite{karjula2026embezzlement,aditya2026nonlocalmagic}. The U(1)-conserving circuit in the Letter realizes a different dynamical mechanism: a fixed global charge, an unequal cut, and a thermodynamic force convert diffusive subsystem-charge fluctuations into a growing width in the logarithmic Schmidt-rank coordinate after the leading entropy has formed. The equilibrium theorem fixes the final-state law; the proposition below states the precise spectral inputs that extend this charge mechanism to a typical circuit trajectory.

For a fixed-charge circuit state, write its reduced density matrix as $\rho_A=\bigoplus_q\rho_{A,q}$, where $\rho_{A,q}$ is the unnormalized block in the eigenspace $Q_A=q$. Its trace $p_q=\Tr\rho_{A,q}$ is the probability of that charge, and the sector dimension is $d_A(q)=\binom{L_A}{q}=a_q$. Define the charge-block-flattened state
\begin{equation}
 \widetilde\rho_A=\bigoplus_q\frac{p_q}{d_A(q)}\mathbb I_{d_A(q)}.
 \label{eq:flattened}
\end{equation}
Equation~\eqref{eq:lipschitz} gives a uniform defect bound in terms of $T(\rho_A,\widetilde\rho_A)$. For the purified operational error, contractivity and the triangle inequality yield
\begin{equation}
 \sup_m|P_m^\star(\rho_A)-P_m^\star(\widetilde\rho_A)|
 \leq2P(\rho_A,\widetilde\rho_A).
 \label{eq:purifiedstability}
\end{equation}
The block-flattened state is a directly computable proxy for the charge envelope, while the information residual gives the corresponding task-specific spectral criterion.

\begin{proposition}[Conditional dynamical law]
Let $[\rho_A,Q_A]=0$, let $\mu\ne0$, and suppose $s=|\mu|\sqrt{\Var Q_A}\to\infty$. Assume that the standardized charge converges to a standard normal variable and that
\begin{equation}
 \frac{K-\langle K\rangle-\mu(Q_A-\langle Q_A\rangle)}{|\mu|\sqrt{\Var Q_A}}
 \xrightarrow{\mathrm{prob.}}0,
 \label{eq:residual}
\end{equation}
where $K=-\ln\lambda$ is sampled with the Schmidt probabilities of the output state and $\xrightarrow{\mathrm{prob.}}$ denotes convergence in probability. Then the Gaussian conversion law, Eq.~\eqref{eq:gausslaw}, holds with scale $s$.
\end{proposition}

\paragraph{Proof.} Equation~\eqref{eq:residual} states that, after centering, $K/s$ differs in probability by a vanishing amount from $\mu Q_A/s$. The charge central limit and Slutsky's theorem give Eq.~\eqref{eq:clt}, after which Theorem~1 applies. A sufficient second-moment condition is $\Var(K-\mu Q_A)=o(\mu^2\Var Q_A)$ by Chebyshev's inequality. Alternatively, the likelihood-ratio argument in Eqs.~\eqref{eq:negativeJ} and \eqref{eq:positiveJ} shows that $D(\rho_A\Vert\ee^{-\mu Q_A}/Z_A)=o(s)$ is sufficient. These conditions allow nontrivial within-sector structure while preserving the same leading information coordinate. \hfill$\square$

For the brick-wall ensemble defined below, Haar averaging in each two-site charge block maps diagonal one-copy populations to the parallel-update symmetric simple exclusion process, in which hard-core particles hop symmetrically between neighboring sites. The resulting charge scale follows the diffusive full-counting-statistics form,
\begin{equation}
 \Var Q_A(t)\sim a\sqrt{D_{\rm diff}t},
 \label{eq:diffusionvariance}
\end{equation}
before finite-size saturation. Here $D_{\rm diff}$ is the diffusion constant and $a$ depends on the initial state, filling, and geometry. When typical circuit trajectories share this charge scale and their information residual is subleading, Eq.~\eqref{eq:gausslaw} gives
\begin{equation}
 B_\epsilon^P(t)\sim\kappa_\epsilon|\mu|\sqrt a\,(D_{\rm diff}t)^{1/4}.
 \label{eq:quarter}
\end{equation}
More generally, suppose uniformly for $\tau=t/L^2$ in a compact interval that $\Var Q_A(t,L)=L\mathcal V(\tau)+o(L)$, the standardized charge obeys a central limit, and Eq.~\eqref{eq:residual} holds. Then
\begin{equation}
 \frac{B_\epsilon^P(t,L)}{\sqrt L}=\kappa_\epsilon|\mu|\sqrt{\mathcal V(\tau)}+o(1).
 \label{eq:fss}
\end{equation}
Here $\kappa_\epsilon=2\sqrt{-\ln(1-\epsilon^2)}/\ln2$. The $t^{1/4}$ law is the hydrodynamic intermediate-time form when $\mathcal V(\tau)\sim a\sqrt{D_{\rm diff}\tau}$ and the logarithmic Schmidt-rank coordinate follows the charge coordinate.

Equations~\eqref{eq:quarter} and \eqref{eq:fss} identify the two physical ingredients: diffusive broadening of the subsystem-charge distribution and the finite-size data scaling.

\section{Charge-adapted two-replica matrix-product state}

The large-$L$ calculation used in the main text averages the two-copy observable at fixed replica number and represents the resulting deterministic tensor network as a symmetry-adapted matrix-product state (MPS), following related treatments of conserved-dynamics observables~\cite{liu2026participation,aditya2026coherence,xiao2026diffusive}. This representation retains the disorder-averaged charge distribution and sector purities needed to connect hydrodynamic broadening with the operational conversion curve, while its symmetry structure makes the large systems in Figs.~2 and 3 accessible.

\subsection{From a random circuit to a deterministic replica transfer matrix}

For one circuit realization $c$, let $\rho_c(t)=|\psi_c(t)\rangle\!\langle\psi_c(t)|$ be the global density matrix. The two-copy object is
\begin{equation}
 \mathcal M_2(t)=\mathbb E_c\bigl[\rho_c(t)^{\otimes2}\bigr].
 \label{eq:M2}
\end{equation}
Vectorizing its two bra indices turns $\mathcal M_2$ into a state on four replica contours. For a two-site gate $U$, the local transfer matrix in ket-pair/bra-pair order is
\begin{equation}
 W_2=\mathbb E_U\left[U^{\otimes2}\otimes U^{*\otimes2}\right].
 \label{eq:W2}
\end{equation}
The gates are independent in space and time, so the full disorder average factorizes: one brick-wall layer of the random circuit becomes one layer of identical deterministic $W_2$ tensors. The replica evolution is therefore deterministic once the Haar average is taken.

The physical two-site gate is block diagonal in the occupation basis $(00,01,10,11)$,
\begin{equation}
 U=U_0\oplus U_1\oplus U_2,
 \qquad U_0,U_2\in U(1),\quad U_1\in U(2),
 \label{eq:u1gate}
\end{equation}
with independent Haar measures on the three blocks, where $U(d)$ denotes the $d$-dimensional unitary group. Haar averaging projects the representation $U^{\otimes2}$ onto its commutant~\cite{weingarten1978,collins2006}, producing the deterministic two-site transfer gate used below.

\subsection{The six local states and what their labels mean}

At each physical site we order the four binary occupations as
\begin{equation}
 \alpha=(k_1,k_2,b_1,b_2),
 \label{eq:replicalabels}
\end{equation}
where $k_1,k_2$ are the two ket-contour occupations and $b_1,b_2$ are the two vectorized bra-contour occupations. These labels refer to the four legs of the two-replica contour. For the diagonal product initial state and the trace/swap observables used below, the transfer network stays in the sitewise neutral subspace
\begin{equation}
 k_1+k_2=b_1+b_2.
 \label{eq:localneutral}
\end{equation}
Its six basis states are
\begin{align}
 &(0,0,0,0),\quad(1,1,1,1),\notag\\
 &(1,0,1,0),\quad(1,0,0,1),\quad(0,1,1,0),\quad(0,1,0,1).
 \label{eq:sixbasis}
\end{align}
Restricting Eq.~\eqref{eq:W2} to this space gives a real nearest-neighbor transfer gate on six local states. This exact neutral-sector reduction is the basis of the charge-adapted MPS calculation.

Equation~\eqref{eq:localneutral} leaves three independent charges. We choose
\begin{equation}
 \bm q_\alpha=(k_1,k_2,b_1),
 \qquad b_2=k_1+k_2-b_1.
 \label{eq:threecharges}
\end{equation}
The transfer gate conserves the sum of each component of $\bm q$ on its two sites. The physical leg therefore carries an exact $U(1)^3$ charge with components $(k_1,k_2,b_1)$~\cite{singh2010,hauschild2018tenpy}. This block structure keeps the replica-charge sectors separate during the MPS evolution and makes the large-system charge and purity contractions used in the main text accessible.

\subsection{Initial state and replica evolution}

The physical circuit starts from the filling-$1/6$ period-6 repeating product state $|100000\,100000\cdots\rangle$. If the physical occupation at site $j$ is $n_j$, the replica MPS starts in the local basis state $(n_j,n_j,n_j,n_j)$. The boundary condition is open, the subsystem is the left block of $L_A=L/3$ sites, and one time step consists of an even-bond layer followed by an odd-bond layer. Each active bond receives the same deterministic gate from Eq.~\eqref{eq:W2}, because the original gates are identically distributed and independently averaged.

The symmetry-resolved MPS retains the replica-charge sectors throughout the evolution, and the observables are normalized with the trace boundary contraction defined below. This structure isolates the charge envelope and the within-sector purity in the same deterministic calculation.

\subsection{Trace, swap, and counting-field contractions}

Let $|\alpha\rangle=|k_1,k_2,b_1,b_2\rangle$ denote one of the states in Eq.~\eqref{eq:sixbasis}. The local trace and replica-swap bras are
\begin{align}
 \langle\mathsf I|&=\sum_{\alpha}\delta_{b_1,k_1}\delta_{b_2,k_2}\langle\alpha|,\notag\\
 \langle\mathbb S|&=\sum_{\alpha}\delta_{b_1,k_2}\delta_{b_2,k_1}\langle\alpha|.
 \label{eq:traceswap}
\end{align}
Contracting $\langle\mathbb S|$ on $A$ and $\langle\mathsf I|$ on its complement gives the annealed subsystem purity,
\begin{equation}
 \frac{\langle\mathbb S_A\mathsf I_B|\mathcal M_2(t)\rangle}
 {\langle\mathsf I_{AB}|\mathcal M_2(t)\rangle}
 =\mathbb E_c\Tr\rho_{A,c}^2.
 \label{eq:totalpurityreplica}
\end{equation}
The denominator supplies the normalization of the replicated observable.

Let $\Pi_q$ denote the projector onto the subsystem-charge sector $Q_A=q$, and define $p_{q,c}=\Tr(\Pi_q\rho_{A,c})$. Charge resolution is obtained by multiplying every trace or swap vector on $A$ by $\exp(i\theta k_1)$. With $\theta_n=2\pi n/(L_A+1)$, the two generating functions are
\begin{align}
 Z_0(\theta)&=\sum_q e^{i\theta q}\,\bar p_q,
 &\bar p_q&=\mathbb E_c p_{q,c},\notag\\
 Z_{\mathbb S}(\theta)&=\sum_q e^{i\theta q}\,R_q,
 &R_q&=\mathbb E_c\Tr\rho_{A,q,c}^2.
 \label{eq:generatingfunctions}
\end{align}
Because $0\leq q\leq L_A$, the discrete Fourier transform on the $L_A+1$ angles is exact. The charge mean and variance used in the main text are evaluated from the normalized $\bar p_q$, and $s_Q=|\mu|\sqrt{\Var_{\bar p}Q_A}$ with $\mu=\ln5$ at filling $1/6$.

Numerically, both generating functions in Eq.~\eqref{eq:generatingfunctions} are divided by the same trace contraction $\mathcal N(t)=\langle\mathsf I_{AB}|\mathcal M_2(t)\rangle$ before the Fourier transform. Thus $\sum_q\bar p_q=1$ and $R_q$ and $\bar p_q^2$ enter the certificate below with consistent normalization.

\subsection{A two-copy bound on the omitted sector structure}

For one circuit realization, define its charge-block-flattened state using its own charge probabilities,
\begin{equation}
 \widetilde\rho_{A,c}=\bigoplus_q\frac{p_{q,c}}{d_A(q)}\mathbb I_{d_A(q)},
 \qquad d_A(q)=\binom{L_A}{q}.
 \label{eq:sampleflattened}
\end{equation}
The trace distance to this state separates into charge sectors. For a traceless Hermitian operator $X$ in dimension $d$, $\|X\|_1\leq\sqrt d\|X\|_2$. Therefore
\begin{align}
 \varepsilon_{{\rm sec},c}
 &\equiv T(\rho_{A,c},\widetilde\rho_{A,c})\notag\\
 &\leq\frac12\sum_q
 \sqrt{d_A(q)\Tr\rho_{A,q,c}^2-p_{q,c}^2}.
 \label{eq:samplecertificate}
\end{align}
Averaging, applying Jensen's inequality sector by sector, and then using $\mathbb E p_{q,c}^2\geq(\mathbb E p_{q,c})^2$ gives the fully two-copy certificate
\begin{align}
 \mathbb E_c\varepsilon_{{\rm sec},c}
 &\leq\frac12\sum_q
 \sqrt{d_A(q)R_q-\mathbb E_c p_{q,c}^2}\notag\\
 &\leq\frac12\sum_q\sqrt{d_A(q)R_q-\bar p_q^2}
 \equiv\varepsilon_{\rm sec}.
 \label{eq:annealedcertificate}
\end{align}
The certificate $\varepsilon_{\rm sec}$ quantifies the spectral correction caused by nonuniformity within the charge sectors. Figure~\ref{fig:replicacontrols}(a) shows that it decreases through the diffusive window, so the charge-flattened spectrum becomes a progressively better description of the sector-resolved output in the regime where the operational growth is observed. At late times each radicand is a small difference between $d_A(q)R_q$ and $\bar p_q^2$, making the numerical value sensitive to MPS truncation and residual normalization error; the endpoint values are conservative order-of-magnitude bounds without bond-dimension extrapolation. Together with Eq.~\eqref{eq:lipschitz}, the certificate gives uniform control on the corresponding majorization defects.

\subsection{The charge-flattened operational proxy}

The mean charge distribution defines a deterministic charge-flattened spectrum,
\begin{equation}
 \overline{\widetilde\rho}_A=
 \bigoplus_q\frac{\bar p_q}{d_A(q)}\mathbb I_{d_A(q)}.
 \label{eq:meanflattened}
\end{equation}
We represent it as a list of eigenvalue--multiplicity pairs $(\bar p_q/d_A(q),d_A(q))$ and apply the exact PAV construction of Sec.~I to their rank boundaries. This sector representation evaluates the operational curve at the scale of the number of charge sectors, while retaining the full degeneracies. We denote the resulting continuously interpolated purified-error yield by $\widetilde B_\epsilon^P$.

The quantity $\widetilde B_\epsilon^P$ is the conversion yield of the deterministic mean charge-flattened spectrum in Eq.~\eqref{eq:meanflattened}. The sector-flatness and charge self-averaging diagnostics below quantify how this charge-only curve organizes the output spectra of individual circuit realizations.

\subsection{Charge self-averaging from virtual-bond sectors}

At the subsystem cut, the two independent ket-contour bond charges are the cumulative subsystem charges $q$ and $r$ of the two replicas. Contracting the identity bra on both sides of that bond and grouping the remaining bond contributions by $(q,r)$ yields
\begin{equation}
 J_{qr}=\mathbb E_c[p_{q,c}p_{r,c}]
 \label{eq:jointpq}
\end{equation}
The same contraction directly gives the joint charge histogram of two replicas. Both factors $p_{q,c}p_{r,c}$ belong to the same circuit realization $c$, so $J_{qr}$ measures circuit-to-circuit fluctuations of the charge distribution. Its marginals reproduce $\bar p_q$. Write $p_c=(p_{q,c})_q$ and $\bar p=(\bar p_q)_q$ for these histograms, with total-variation distance $T(p_c,\bar p)=\frac12\sum_q|p_{q,c}-\bar p_q|$. The diagonal entries give $\Var_c(p_{q,c})=J_{qq}-\bar p_q^2$, and Cauchy--Schwarz therefore bounds the mean fluctuation by
\begin{equation}
 \mathbb E_c T(p_c,\bar p)
 \leq\frac12\sum_q\sqrt{J_{qq}-\bar p_q^2}.
 \label{eq:tvselfaveraging}
\end{equation}
From the full joint distribution, the disorder variance of the trajectory-dependent mean charge is
\begin{equation}
 \Var_c\!\left(\sum_q q p_{q,c}\right)
 =\sum_{q,r}qrJ_{qr}-\left(\sum_q q\bar p_q\right)^2.
 \label{eq:disordervariance}
\end{equation}
The total variance of the mean distribution separates exactly as
\begin{equation}
 \Var_{\bar p}(Q_A)
 =\mathbb E_c\Var_{p_c}(Q_A)
 +\Var_c\!\left(\sum_q q p_{q,c}\right).
 \label{eq:totalvariancedecomposition}
\end{equation}
The circuit-to-circuit contribution to the charge variance is small compared with the total variance, and the total-variation bound is small at the matched large-system time. Figure~\ref{fig:replicacontrols}(b) therefore supports the interpretation that the broad charge envelope reflects collective transport, with drift of the circuit-averaged mean remaining subleading.

Figure~\ref{fig:replicacontrols}(a) shows the sector-flatness certificate decreasing through the diffusive window, while Fig.~\ref{fig:replicacontrols}(b) shows that circuit-to-circuit fluctuations of the charge envelope remain small. Together, these observations support a physical separation between collective charge transport, which sets the broad operational coordinate, and within-sector scrambling, which brings the output spectrum toward the charge-flattened form.

\section{Fixed-transport comparison}

In addition to the charge-adapted replica-MPS calculation, we directly evolve the finite-size full state with TensorCircuit-NG and TenCirPauli~\cite{zhang2023tensorcircuit,zhang2026tensorcircuitng} and construct its reduced density matrix. This gives direct access to the charge-sector weights, within-sector spectra, and operational conversion curve. We use this full-state calculation as a finite-size test.

The main dynamical proxy combines two physical processes: charge redistribution across sectors and scrambling within each sector. To test whether the operational curve is primarily organized by the charge coordinate, we change a genuine interaction phase while preserving the same annealed one-copy population map and compare the resulting conversion curves at matched charge widths.

To vary internal mixing without changing the averaged one-copy charge map, each two-site gate is constructed as
\begin{equation}
 U_\gamma=1\oplus V\oplus\left[\det(V)\ee^{i\gamma\xi}\right],
 \qquad \xi\sim\operatorname{Uniform}[-\pi,\pi],
 \label{eq:intervention}
\end{equation}
where $V$ is Haar random in $U(2)$. The physically relevant two-particle interaction phase is the gauge-invariant combination
\begin{equation}
 \Delta=\phi_{00}+\phi_{11}-\arg\det V=\gamma\xi \pmod{2\pi}.
 \label{eq:interactionphase}
\end{equation}
At $\gamma=1$ this is the original independent Haar $U(1)\times U(2)\times U(1)$ ensemble up to an irrelevant gate-global phase, while $\gamma=0$ is the number-conserving matchgate (free-fermion) boundary. We use $\gamma=1/2$ as an intermediate interaction-phase deformation: all values have the same annealed charge-transport map, but differ in within-sector scrambling.

\begin{figure}[t]
\centering
\includegraphics[width=0.98\textwidth]{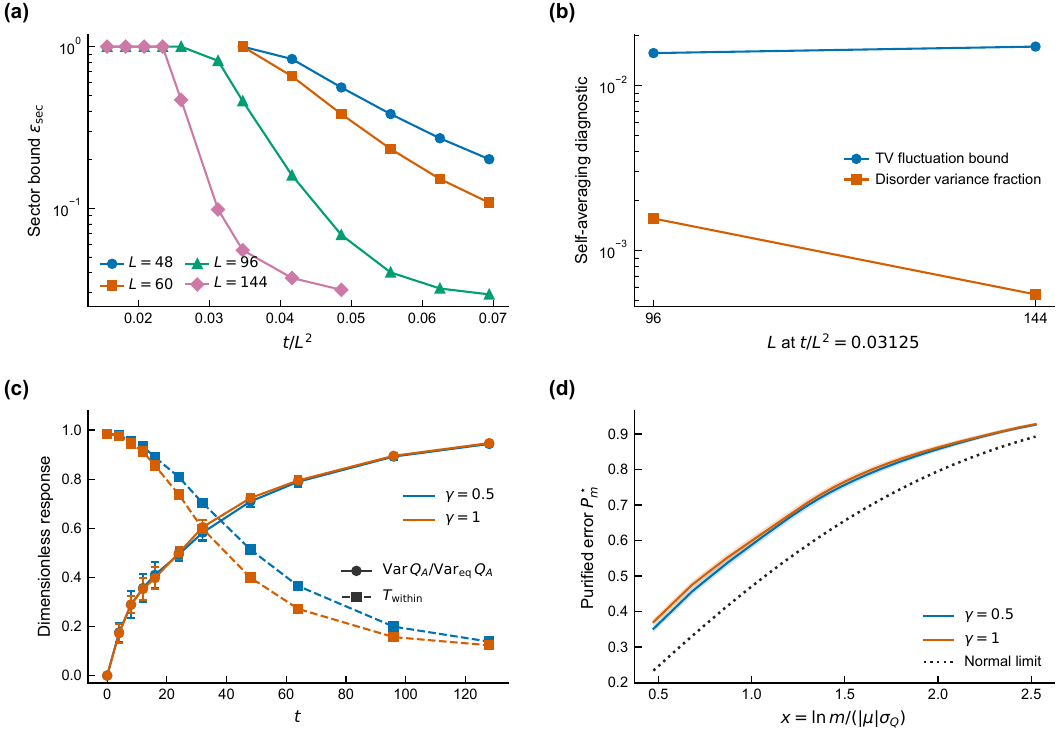}
\caption{\label{fig:replicacontrols}Charge self-averaging and sector mixing in the dynamical calculation. (a) The two-copy sector-flatness certificate $\varepsilon_{\rm sec}$ decreases through the diffusive window for $L=48,60,96,144$, showing that the charge-flattened spectrum increasingly captures the sector-resolved output. (b) At $t/L^2=0.03125$, the upper bound on the total-variation distance $T(p_c,\bar p)$ and the fraction of $\Var_{\bar p}(Q_A)$ generated by circuit-to-circuit motion of $\langle Q_A\rangle_c$ are both small, showing that the broad charge envelope is self-averaging and transport dominated. (c) Fixed-transport comparison at $L=36$: charge variance normalized by its equilibrium value (solid circles) and within-sector trace distance (dashed squares) for $\gamma=0.5$ and $1$. Points and error bars are means and standard errors over four paired realizations. The charge curves nearly coincide while sector mixing separates. (d) At $L=36,t=96$, exact mean purified-error curves with standard-error bands for the two numerical setups $\gamma=0.5$ and $1$, together with the theoretical normal-limit curve. The paired root-mean-square (RMS) distance between the $\gamma=0.5$ and $\gamma=1$ curves is $0.01100\pm0.00245$; separately, their RMS distances from the normal-limit curve are $0.0952$ and $0.1050$, respectively.}
\end{figure}

To quantify within-sector mixing, we use the average entropy within charge sectors, $S_{\rm cond}=\sum_qp_qS(\rho_{A,q}/p_q)$, where $S(\sigma)=-\Tr(\sigma\ln\sigma)$, and the within-sector trace distance
\begin{equation}
 T_{\rm within}=T(\rho_A,\widetilde\rho_A)
 =\frac12\sum_q\left\|\rho_{A,q}-\frac{p_q}{d_A(q)}\mathbb I_{d_A(q)}\right\|_1.
 \label{eq:withinsector}
\end{equation}
We compare the ensembles at $L=36,t=96$, where the mean entropy is close to equilibrium while the charge variance remains below its equilibrium value. Within each of four paired realizations, the same $V$ and $\xi$ are used and only $\gamma$ is changed. The two full-state ensembles have nearly identical charge variance and charge-rescaled operational curves, while their within-sector entropy and trace-distance diagnostics differ modestly; these comparisons are shown in Fig.~\ref{fig:replicacontrols}(c,d). Here $\sigma_Q^2=\Var(Q_A)$ for the corresponding output state, and the operational curves are compared as functions of the charge-scaled target coordinate $x=\ln m/(|\mu|\sigma_Q)$.

Figure~\ref{fig:replicacontrols}(c) shows that the normalized charge-variance curves for $\gamma=1/2$ and $1$ remain nearly coincident. Panel (d) shows that the two charge-rescaled operational curves differ by only $0.01100\pm0.00245$ in paired RMS distance. The intervention therefore supports charge width as the leading dynamical coordinate and identifies within-sector scrambling as a subleading finite-size correction in this regime.

\clearpage
\bibliography{ref}